\documentclass[10pt,final,journal,twocolumn]{IEEEtran}
\usepackage{cite}
\usepackage[cmex10]{amsmath}
\usepackage{placeins}
\usepackage{graphicx,epsfig,amssymb}
\usepackage{anysize}
\marginsize{0.7in}{0.7in}{0.8in}{1.2in} 

\begin{document}

\title{Spatial Spectrum and Energy Efficiency of Random Cellular Networks}
\author{Xiaohu~Ge,~\IEEEmembership{Senior~Member,~IEEE,}
        Bin~Yang,~\IEEEmembership{Student~Member,~IEEE,}
        Junliang~Ye,
        Guoqiang~Mao,~\IEEEmembership{Senior~Member,~IEEE,}
        Cheng-Xiang~Wang,~\IEEEmembership{Senior~Member,~IEEE,}
        Tao~Han,~\IEEEmembership{Member,~IEEE}

\thanks{\scriptsize{Manuscript received July 09, 2014; revised October 19, 2014; accepted January 10, 2015. The Associate Editor for this paper was O. Oyman. The authors would like to acknowledge the support from the International Science and Technology Cooperation Program of China under the grants 2014DFA11640 and 0903, the National Natural Science Foundation of China (NSFC) under the grants 60872007, 61271224 and 61471180, NFSC Major International Joint Research Project under the grant 61210002, the Fundamental Research Funds for the Central Universities under the grants 2013ZZGH009 and 2014QN155. This research is partially supported by EU FP7-PEOPLE-IRSES, project acronym S2EuNet (grant no. 247083), project acronym WiNDOW (grant no. 318992) and project acronym CROWN (grant no. 610524). The work of G. Mao was supported by the Australian Research Council Discovery projects DP110100538 and DP120102030. C.-X. Wang acknowledges the support of his work by the 863 project in 5G wireless networking, Ministry of Science and Technology of China (Grant No. 2014AA01A701), the Opening Project of the Key Laboratory of Cognitive Radio and Information Processing (Guilin University of Electronic Technology), Ministry of Education (No. 2013KF01), and the EU FP7 QUICK project (Grant No. PIRSES-GA-2013-612652). \emph{(Corresponding author: T. Han.)}}}
\thanks{\scriptsize{X. Ge, B. Yang, J. Ye and T. Han are with the School of Electronic Information and Communications, Huazhong University of Science and Technology, Wuhan 430074, Hubei, China (email: \{xhge, yangbin, yejunliang, hantao\}@mail.hust.edu.cn).}}
\thanks{\scriptsize{G. Mao is with  University of Technology Sydney and National ICT Australia, Sydney, Australia (email: guoqiang.mao@uts.edu.au).}}
\thanks{\scriptsize{C.-X. Wang is with Institute of Sensors, Signals and Systems, School of Engineering \& Physical Sciences, Heriot-Watt University, Edinburgh, EH14 4AS, UK (email: cheng-xiang.wang@hw.ac.uk).}}}
\maketitle

\markboth{IEEE Trans. on COMMUNICATIONS, Vol. XX, No. Y, Month 2015} {Ge \textit{et al.}:~ Spectrum and Energy Efficiency of Cellular Networks \ldots}%
\begin{abstract}

It is a great challenge to evaluate the network performance of cellular mobile communication systems. In this paper, we propose new spatial spectrum and energy efficiency models for Poisson-Voronoi tessellation (PVT) random cellular networks. To evaluate the user access the network, a Markov chain based wireless channel access model is first proposed for PVT random cellular networks. On that basis, the outage probability and blocking probability of PVT random cellular networks are derived, which can be computed numerically. Furthermore, taking into account the call arrival rate, the path loss exponent and the base station (BS) density in random cellular networks, spatial spectrum and energy efficiency models are proposed and analyzed for PVT random cellular networks. Numerical simulations are conducted to evaluate the network spectrum and energy efficiency in PVT random cellular networks.

\end{abstract}

\section{Introduction}
\label{sec1}

 \IEEEPARstart{W}{ith} the development of the new generation cellular networks, more than 2,000,000 base stations (BSs) have been deployed until 2012. Report from the US shows that the energy consumption of an on-grid BS costs 3,000 dollars per year on the average while the average energy bill for running an off-grid BS reaches 30,000 dollars per year \cite{Hasan11}. Considering above energy cost and the carbon footprint in telecommunication systems, energy efficiency of cellular networks has received much attention from service providers \cite{Fehske11}. Therefore, it is important to investigate the spectrum and energy efficiency of cellular mobile communication systems \cite{Ku13Spectral, Hong13Energy, Ge14Energy}.


In cellular networks, a significant amount of power is consumed for transmitting signal over wireless channels. Therefore, the wireless channel model is a key metric for the performance evaluation of cellular networks. Gilbert-Elliott channel models were recommended as the most typical discrete channel model, where channel states are divided into the ¡°good¡± or ¡°bad¡± state \cite{Gilbert60}. Based on the Gilbert-Elliott channel model, the network throughput and the bit error rate were investigated for wireless networks with bursty and noisy channels \cite{Elliott63}. Furthermore, the delay and throughput of cellular networks were analyzed based on Gilbert-Elliott channel models \cite{Jayaparvathy05,Anand03}. Recently, finite state Markov modeling of fading channel was extended from wired network to wireless networks and used for wireless network performance analysis \cite{Sadeghi08}. Particularly, the received signal-to-noise ratio (SNR) was partitioned into a finite number of states and then a finite state Markov chain (FSMC) was constructed for modeling Rayleigh fading channels \cite{Park09}. Both numerical and simulation results have shown that the FSMC channel model provides an accurate model for the real channels. Based on a three-dimensional Markov chain model, a pico-cellular airport traffic model was proposed in \cite{Bhattacharya10} which models Engset distributed fresh call arrival process and the general distributed handoff process with dynamic channel allocation. The energy efficiency of fixed-rate transmissions from Markov sources over Rayleigh fading channels was investigated in \cite{Ozmen14}. On that basis, a closed-form expression of the minimum energy per bit was derived. Using the Markov chain to model the channel access in femtocell networks, energy and spectrum efficiency models were proposed in \cite{Ge14} for a two-tier femtocell network with partially open channels.

Although one-dimensional linear model and two-dimensional lattices model including square lattices and triangular lattices were widely used in the modeling of cellular networks \cite{Lee86}, the assumption of regular deployment of BSs in a plane is not consistent with real BSs deployment. Ignoring these structural fluctuations of the geometric objects may cause significant bias on the evaluation of the cellular network performance \cite{Baccelli97}. To evaluate the geometric structure of BSs coverage, the Poisson point process has been presented for modeling of BSs spatial structure \cite{Andrews11,Guo13}. Using realistic BSs locations collected from the Ofcom region UK, the Poisson point process has been justified for modeling of BS locations in cellular networks \cite{Guo13}. Moreover, in many cellular network scenarios only a statistical description of BSs locations is available. For these scenarios, the stochastic geometric model provides a promising solution for modeling and performance analysis of cellular networks. Particularly, in \cite{Win09}, Win {\em et al.} introduced a mathematical framework for the characterization of network interference in which the interferers are scattered according to a spatial Poisson point process and are operating asynchronously in a wireless environment subject to path loss, shadowing and multipath fading. Using physically realistic stochastic models for BS's locations in multi-tier heterogeneous networks, a general signal-to-interference-and-noise ratio (SINR) distribution was derived in \cite{Mukherjee12} for calculating the probability of the user being able to camp on a macrocell or an open-access femtocell in the presence of closed subscribe group femtocells. Assuming mobile users and BSs in different tiers of heterogeneous cellular networks are independent Poisson point processes, a tractable model of downlink heterogeneous cellular network consisting of K tiers of randomly located BSs was developed in \cite{Dhillon12}. By assuming spatial Poisson distribution of transmitters in the primary network, the achievable transmission capacity of a secondary cognitive mesh network sharing the uplink spectrums with cellular users under the outage probability constraints of both the primary and secondary systems was investigated in \cite{Jing12}. Using a stochastic geometry based model and different sleeping policies, the success probability and energy efficiency in homogeneous macrocell and heterogeneous K-tier wireless network were derived in \cite{Soh13}. In \cite{Deng12}, based on the spatial Poisson point process, the distribution of SINRs and mean achievable rates of both non-cooperative users and cooperative users were derived to evaluate the energy efficiency of relay-assisted cellular networks.


In all aforementioned studies, Markov chain models and stochastic geometry tools are successfully used for investigating cellular networks, respectively. However, the study combining the Markov chain model with random cellular networks is surprisingly lacking in the open literatures. We started an initial study and published the initial result in \cite{Xiaohu14}. One of the major obstacles in combining the Markov chain model with random cellular networks is how to model the user access in stochastic geometry cellular scenarios by Markov chains. To overcome this obstacle, we propose a Markov chain based channel access model for a Poisson-Voronoi tessellation (PVT) random cellular scenario. Furthermore, the Markov chain based channel access model for one PVT cell has been extended to the whole PVT random cellular network using the Palm theory. As a consequence, the spatial spectrum and energy efficiencies of PVT random cellular networks has been analyzed in this paper. The contributions and novelties of this paper are summarized as follows.

\begin{enumerate}
\item Although the Markov chain model and the stochastic geometry theory have been widely used in cellular networks, it is still a great challenge to integrate the Markov chain model into the analysis of stochastic geometry cellular networks. Based on the PVT cellular scenario, a Markov chain based channel access model is first proposed and integrated into random cellular networks.
\item From the proposed wireless channel access model, we derive the exact outage probability and blocking probability for PVT random cellular networks with a spatial Poisson distribution of interfering transmitters, taking into account fading and shadowing effects over wireless channels.
\item Using the outage probability and blocking probability, a new spatial spectrum and energy efficiency model of PVT random cellular networks is developed for performance evaluation.
\item We study the spatial spectrum and energy efficiency of PVT random cellular networks in details and present some interesting observations.
\end{enumerate}

The remainder of this paper is outlined as follows. Section~\ref{sec2} describes the system model. In Section~\ref{sec3}, a wireless channel access model based on Markov chain is proposed and on that basis the outage probability and blocking probability of PVT random cellular networks are derived. Furthermore, the spatial spectrum and energy efficiency models of PVT random cellular networks are presented and evaluated in Section~\ref{sec4}. Finally, Section~\ref{sec5} concludes this paper.

\section{System Model}
\label{sec2}

Assume that both mobile users (MUs) and BSs are located randomly in the infinite plane $\mathbb{R}^2$. Moreover, locations of MUs and BSs are modeled as two independent Poisson point processes, which are denoted as ${\Theta _U}{\rm{ = }}\left\{ {{x_i}:i = 0,1,2, \ldots } \right\}$  and ${\Theta _B} = \left\{ {{y_j},j = 0,1,2, \ldots } \right\}$, where ${x_i}$ and ${y_j}$ are two-dimensional Cartesian coordinates, denoting locations of the ${i}$th MS ${MU_i}$ and the ${j}$th BS ${BS_j}$, respectively. The corresponding intensities of the two Poisson point processes are ${\lambda _U}$ and ${\lambda _B}$, respectively. For a cell, the interference among MUs is ignored and the co-channel interference is assumed to be mainly contributed by adjacent BSs.

\subsection{Wireless Propagation Environments}
For an MU ${x_i} \in {\Theta _U}$ and the associated BS ${y_j} \in {\Theta _B}$, the channel gain ${L_{{y_j}}}({x_i})$, defined as the ratio between the received power ${P_{r\_{x_i}}}$ at an MU and the corresponding transmission power ${P_{{y_j}}}$ from its associated BS, is \[{L_{{y_j}}}({x_i}) = \frac{{{P_{r\_{x_i}}}}}{{{P_{{y_j}}}}} = \frac{{K \cdot {S_{{y_j}}}({x_i})}}{{L\left( {\left\| {{y_j} - {x_i}} \right\|} \right)}}{\rm{ = }}\frac{{K \cdot \prod\nolimits_k {Z_{{y_j}}^k({x_i})} }}{{{{\left\| {{y_j} - {x_i}} \right\|}^b}}},\tag{1}\]
where ${K}$ is a constant depending on antenna gains, ${S_{{y_j}}}({x_i}){\rm{ = }}\prod\nolimits_k {Z_{{y_j}}^k({x_i})}$ is the impact of fading and shadowing effects on the received signal power in wireless propagation environments; $\left\{ {\left. {Z_{{y_j}}^k({x_i})} \right|k \in {\mathbb{Z}^ + }} \right\}$ are independent random variables (RVs), which account for propagation effects such as fading and shadowing, where $\mathbb{Z}^ +$ is the positive integer set, the values of ${k}$ and the distribution of $\left\{ {Z_{{y_j}}^k({x_i})} \right\}$ are general enough to account for various propagation scenarios, including the following:
\begin{enumerate}
\item ${S_{{y_j}}}({x_i}) = Z_{{y_j}}^1({x_i}) = 1$: without fading and shadowing effects;
\item $Z_{{y_j}}^1({x_i})$ follows an exponential distribution: Rayleigh fading propagation scenario;
\item ${S_{{y_j}}}({x_i}) = Z_{{y_j}}^1({x_i}) = {\varsigma ^2}$ with ${\varsigma ^2} \sim \Gamma \left( {m,\frac{1}{m}} \right)$, where $\Gamma \left( {m,\frac{1}{m}} \right)$ is a Gamma distribution with 1 mean and variance of $\frac{1}{m}$: Nakagami-m fading propagation scenario;
\item ${S_{{y_j}}}({x_i}) = Z_{{y_j}}^2({x_i}) = {e^{2\sigma G}}$ with $G \sim Gaussian\left( {0,1} \right)$, where $\sigma$ is the shadowing coefficient: log-normal shadowing propagation scenario;
\item ${S_{{y_j}}}({x_i}) = \prod\nolimits_k {Z_{{y_j}}^k({x_i})}  = Z_{{y_j}}^1({x_i})Z_{{y_j}}^2({x_i}) = {\varsigma ^2} \cdot {e^{2\sigma G}}$: Nakagami-m fading and log-normal shadowing propagation scenario.
\end{enumerate}
$L\left( {\left\| {{y_j} - {x_i}} \right\|} \right)$ accounts for the far-field path loss, which is a function of the distance between the MU located at ${x_i}$ and the BS located at ${y_i}$, denoted as ${\left\| {{y_j} - {x_i}} \right\|^b}$, where $b$ is the path loss exponent. In general, the path loss exponent is environment-dependent and can approximately range from 0.8 (e.g., hallways inside buildings) to 6 (e.g., buildings without line of sight) \cite{Win09,Rappaport96}.

\subsection{User Association Scheme}
\begin{figure}
\centerline{\includegraphics[width=8cm,height=8cm,draft=false]{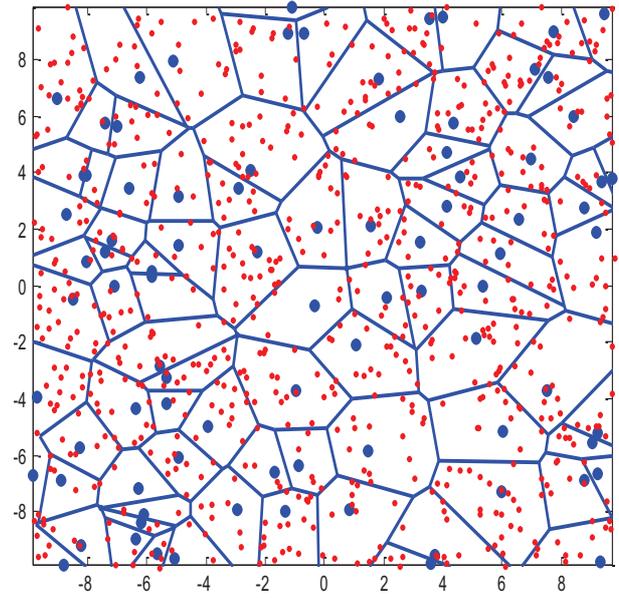}}
\caption{\small An illustration of the PVT random cellular network: blue points are BSs and red points are MUs.}
\label{fig1}
\end{figure}

In realistic cellular networks, the MU receives different pilot signals from adjacent BSs and then associates with a BS based on a specified user association scheme. In recent studies, different user association schemes were adopted for evaluating the performance of cellular networks, for example, the nearest BS association scheme was adopted in \cite{Mukherjee12,Soh13,Qualcomm10}, the highest SINR association scheme was adopted in \cite{Dhillon12,Novlan11,Dhillon11}, and the maximum received signal power association scheme was adopted in \cite{Ali11,Jo12,deLima13}. When an MU located at ${x_i}$ is associated with a BS located at $y_j$, a general association scheme is expressed as \cite{Fooladivanda13,Singh13}.
\[y_j^* = \arg \mathop {\max }\limits_{{y_j} \in {\Theta _B}} {{\rm T}_{{y_j}}}{\left\| {{y_j} - {x_i}} \right\|^{ - b}},\tag{2}\]
where ${{\rm T}_{{y_j}}}$ is the weight association metric, which is dependent on the particular user association scheme. In this paper, the MU is assumed to be associated with the nearest BS in the plane. In this case, the weight association metric is configured as ${{\rm T}_{{y_j}}} = 1$. Accordingly, (2) can be rewritten as
\[y_j^* = \arg \mathop {\max }\limits_{{y_j} \in {\Theta _B}} {\left\| {{y_j} - {x_i}} \right\|^{ - b}}.\tag{3}\]
Based on the PVT method, the coverage of BS located at ${y_j}$ is defined as \cite{Singh13}
\begin{equation}
\begin{aligned}
{\mathcal{C}_{{y_j}}} = \{&y \in \mathbb{R}^2:\left\| {y - {y_j}} \right\| \leqslant \left\| {y - {y_l}} \right\|,\forall {y_l} \in {\Theta _B}
\:and\: \\
& {y_l} \ne {y_j} \}.
\end{aligned}
\label{eq4}
\tag{4}
\end{equation}
Based on (\ref{eq4}), an illustration of the PVT random cellular network is depicted in Fig.~\ref{fig1}, where each cell is denoted as ${{\cal C}_{{y_j}}}$. In Fig.~\ref{fig1}, the cells are rounded by blue lines, the BSs are denoted by blue points and the MUs are denoted by red points. Such stochastic and irregular-topology network forms the so-called PVT random cellular network.

Despite of its apparent complexity, an outstanding property of PVT random cellular networks is that the geometric characteristics of any cell ${{\cal C}_{{y_j}}}$ coincide with that of a typical PVT cell ${{\cal C}_{ori}}$ where the corresponding BS is located at a fixed position, e.g., the origin, according to the Palm theory \cite{Xiang13,Stoyan96}. This feature implies that the analytical results obtained for a typical PVT cell ${{\cal C}_{ori}}$ can be extended to the whole PVT random cellular network. Without loss of generality, the coverage of the typical cell ${{\cal C}_{ori}}$ is expressed as
\begin{equation}
\begin{aligned}
{\mathcal{C}_{ori}} = \{&y \in \mathbb{R}^2:\left\| {y - ori} \right\| \leqslant \left\| {y - {y_l}} \right\|,\forall {y_l} \in {\Theta _B}
\:and\: \\
& {y_l} \ne ori \},
\end{aligned}
\label{eq5}
\tag{5}
\end{equation}
where $ori = (0,0)$ is the origin point in the plane $\mathbb{R}^2$.

\subsection{Wireless Channel Allocation Strategy}
In this paper, a centralized channel allocation strategy is used for PVT random cellular networks. The traffic channels are allocated for a MU after this MU is associated with a specific BS. Furthermore, every MU is assumed to have the perfect channel state information (CSI) of all channels available to its BS. This means the associated MU can get the SINR over all channels. When the SINR value of a channel is larger than or equal to the threshold ${\gamma _0}$, this channel is available for the associated MU and will be marked as 1. When the SINR value of a channel is less than the threshold ${\gamma _0}$, this channel is unavailable for the associated MU and will be marked as 0. All available channel states will be feedbacked to the control centres of PVT random cellular networks, and then the control centres will allocate BS channels to all associated MUs based on the set of available channels.

\section{Models of PVT Random Cellular Networks}
\label{sec3}
\subsection{Markov Chain Model of PVT Random Cellular Networks}
Without loss of generality, a typical cell ${{\cal C}_{ori}}$ in the PVT cellular network is selected for Markov chain based modeling and performance analysis in this paper. The arrival rate of calls to the system is assumed to follow a Poisson distribution with mean ${\lambda}$. The user session duration ${T_S}$ and the cell dwelling time ${T_D}$ are assumed to be governed by exponential distributions with mean ${\mu}$ and ${1 \mathord{\left/
 {\vphantom {1 {{\mathbb{T}_D}}}} \right.
 \kern-\nulldelimiterspace} {{\mathbb{T}_D}}}$ respectively, where ${{\mathbb{T}_D}}$ is the mean cell dwelling time. The channel holding time ${T_H}$ is the minimum of the user session duration and the cell dwelling time, i.e., ${T_H} = \min \left( {{T_S},{T_D}} \right)$, with mean $\eta  = \mu  + {1 \mathord{\left/ {\vphantom {1 {{\mathbb{T}_D}}}} \right. \kern-\nulldelimiterspace} {{\mathbb{T}_D}}}$ . Unlike the discrete Gilbert-Elliott channel model used in many articles \cite{Gilbert60,Elliott63}, in this paper the continuous Gilbert-Elliott channel model is used to model the transitions between the available channels and the unavailable channels. For traditional Markov chain models of wireless channel access in cellular networks, the state change of unavailable BS channels has been modeled by a Markov chain and the state dwelling period of unavailable BS channels is governed by an exponent distribution \cite{Jayaparvathy05,Anand03}. Considering that the number of available BS channel is determined by the total BS channel number minus the unavailable BS channel number in a cell, the state dwelling period of available BS channels is also assumed to be governed by an exponent distribution. The transition rate from the unavailable channel to the available channel is denoted as ${\alpha}$ and the transition rate from the available channel to the unavailable channel is denoted as ${\beta}$. The two parameters will be determined later.

\begin{figure}
\centerline{\includegraphics[width=8.2cm,draft=false]{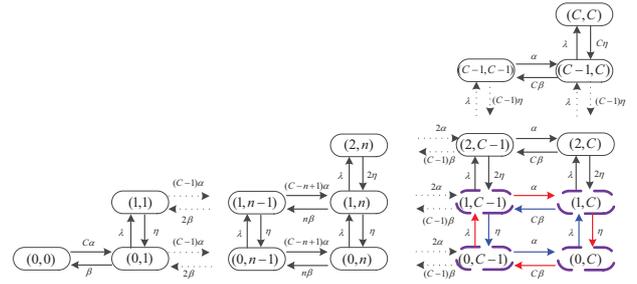}}
\caption{\small Markov chain transition diagram.}
\label{fig2}
\end{figure}

 A two-tuple of nonnegative integers $\left( {m,n} \right)$ is used to denote the network states, where $m$ is the number of occupied channels and $n$ is the total number of channels available to be allocated for MUs in the typical cell ${{\cal C}_{ori}}$, including the channels that have already been allocated (i.e. occupied). Let $C$ be the maximum available channels in the typical cell ${{\cal C}_{ori}}$. Obviously, the constraint $m \le n \le C$ has to be observed. The Markov Chain transition diagram is illustrated in Fig.~\ref{fig2}.

The Markov Chain state transitions in Fig.~\ref{fig2} are described as follows:
\begin{enumerate}
\item $\left( {m,n} \right) \to \left( {m + 1,n} \right)$: When a new call arrives, the number of occupied channels is increased by 1 if the number of occupied channels is less than the number of available channels, i.e., $m < n$.
\item $\left( {m,n} \right) \to \left( {m - 1,n} \right)$: When a call has been successfully serviced, the number of occupied channels is reduced by 1 if this occupied channel is released.
\item $\left( {m,n} \right) \to \left( {m,n + 1} \right)$: An unavailable channel becomes available to be allocated for mobile users due to the time-varying interference, thus the number of available channels is increased by 1.
\item $\left( {m,n} \right) \to \left( {m,n - 1} \right)$: An available channel becomes unavailable to be allocated for mobile users due to the time-varying interference, thus the number of available channels is reduced by 1.
\end{enumerate}

Based on the Kolmogorov¡¯s criteria \cite{Kelly79}, a Markov process is reversible and its stationary state distribution exists if its transition rates satisfy
\[\begin{array}{l}\label{eq6}
\Lambda \left( {{S_1},{S_2}} \right) \cdot \Lambda \left( {{S_2},{S_3}} \right) \cdot  \ldots  \cdot \Lambda \left( {{S_{u - 1}},{S_u}} \right) \cdot \Lambda \left( {{S_u},{S_1}} \right)\\
 = \Lambda \left( {{S_1},{S_u}} \right) \cdot \Lambda \left( {{S_u},{S_{u - 1}}} \right) \cdot  \ldots  \cdot \Lambda \left( {{S_3},{S_2}} \right) \cdot \Lambda \left( {{S_2},{S_1}} \right)
\end{array}\tag{6}\]
for any finite sequence of states ${S_1},{S_2}, \cdots ,{S_u} \in \Xi $, where $\Xi $ is the set of all possible states in this Markov process and $\Lambda \left( {{S_h},{S_k}} \right)$ denotes the transition rate from state $S_h$ to state $S_k$. It is observed that states in Fig.~\ref{fig2} meet (\ref{eq6}) and the stationary state distribution exists as a result. For an easy illustration, four states illustrated with purple dash line are taken as an example. State transition rates on the left and right side of (\ref{eq6}) are placed beside the red arrows and blue arrows, respectively. In this way, the product term on the left side of (\ref{eq6}) is $\alpha  \cdot \eta  \cdot C\beta  \cdot \lambda $ and $\eta \cdot \alpha \cdot \lambda \cdot C\beta $ on the right side. Obviously, $\alpha  \cdot \eta  \cdot C\beta  \cdot \lambda = \eta \cdot \alpha \cdot \lambda \cdot C\beta $. Similarly, we can verify other finite sequences of states and find that (\ref{eq6}) holds. According to the Markov chain state transitions in Fig.~\ref{fig2}, global equilibrium equations are listed as (\ref{eq7}).
\begin{figure*}[!t]
\begin{equation}\label{eq7}
 \left\{ \begin{gathered}
   C\alpha\cdot\pi (0,0) = \beta\cdot\pi (0,1) \hfill \\
  [\beta  + \lambda  + (C - n)]\cdot\pi (0,n) = (C - n + 1)\alpha\cdot\pi (0,n - 1) + \eta\cdot\pi (1,n) + (n + 1)\beta \pi (0,n + 1)\quad  \hfill \\
  \quad for\;0 < n < C \hfill \\
  [C\beta  + \lambda ]\cdot\pi (0,C) = \alpha\cdot\pi (0,C - 1) + \eta\cdot\pi (0,C) \hfill \\
  [C\beta  + \lambda  + m\eta ]\cdot\pi (1,C) = \alpha\cdot\pi (m,C - 1) + (m + 1)\eta\cdot\pi (m + 1,C) + \lambda\cdot\pi (m - 1,C)\quad  \hfill \\
   \quad for\;0 < m < C \hfill \\
  C\eta\cdot\pi (C,C) = \lambda\cdot\pi (C - 1,C) \hfill \\
  [(C - m)\alpha  + m\eta ]\cdot\pi (m,m) = \lambda\cdot\pi (m - 1,m) + (m + 1)\beta\cdot\pi (m,m + 1)\quad  \hfill \\
  \quad for\quad 0 < m = n < C \hfill \\
  [n\beta  + (C - n)\alpha  + \lambda  + m\eta ]\cdot\pi (m,n) = (C + 1 - n)\alpha\cdot\pi (m,n - 1) + (n + 1)\beta\cdot\pi (m,n + 1) +  \hfill \\
  \quad (m + 1)\eta\cdot\pi (m + 1,n) + \lambda\cdot\pi (m - 1,n)\quad for\;0 < m < n < C \hfill \\
\end{gathered}.  \right.
\tag{7}
\end{equation}
\end{figure*}
Based on the queuing theory, the stationary state probabilities are derived as follows
\begin{equation}\label{8}
 \pi \left( {m,n} \right) = \left\{ \begin{gathered}
  \frac{1}{\chi }{\left( {\frac{\lambda }{\eta }} \right)^m}\frac{1}{{m!}} {\binom{C}{n}} {\left( {\frac{\alpha }{\beta }} \right)^n} \hfill \\
  \chi  = \sum\limits_{m \leqslant n \leqslant C} {{{\left( {\frac{\lambda }{\eta }} \right)}^m}\frac{1}{{m!}}{\binom{C}{n}}{{\left( {\frac{\alpha }{\beta }} \right)}^n}}  \hfill \\
\end{gathered}  \right.,\tag{8}
\end{equation}
where $\binom{C}{n}$ is the binomial coefficients meaning the number of ways of picking $n$ unordered outcomes from $C$ possibilities.

Based on the Gilbert-Elliott channel model \cite{Jayaparvathy05,Anand03}, the probability of a channel being unavailable $\varepsilon $ is defined as
\begin{equation}\label{9}
\varepsilon  = \frac{\beta }{{\alpha  + \beta }}.\tag{9}
\end{equation}
The ratio between ${\alpha}$ and ${\beta}$ needs to be calculated in (\ref{8}). Based on (\ref{9}), $\alpha /\beta $ equals to $\left( {1 - \varepsilon } \right)/\varepsilon $. In this paper, the channel is unavailable when the SINR value of channel is less than the given threshold. In this case, the probability of unavailable channel is equal to the expected value of the outage probability with different number of interferers. $\varepsilon $ then is expressed as follows
\begin{equation}\label{10}
\varepsilon  = \mathop {\lim }\limits_{{N_I} \to \infty } \sum\limits_{\vartriangle  = 1}^{{N_I}} {{p_{out}}\left( {{\gamma _0},\Delta } \right){\binom{N_I}{\Delta}}{p^\Delta }{{\left( {1 - p} \right)}^{{N_I} - \Delta }}}, \tag{10}
\end{equation}
where $N_I$ is the maximum number of interferers in ${\Theta _B}$; $\Delta (\Delta  \le {N_I})$ is the possible number of interferers aggregated at a wireless channel; ${p_{out}}\left( {{\gamma _0},\Delta } \right)$ is the outage probability conditioned on that there are ${\Delta}$ interferers; $p$ is the probability that a channel is being used at present, i.e.,
\begin{equation}\label{11}
p = \sum\limits_{m = 0}^n {\sum\limits_{n = 0}^C {\frac{m}{C}\pi \left( {m,n} \right)} }. \tag{11}
\end{equation}

It is observed that the stationary state probability ${\alpha  \mathord{\left/
 {\vphantom {\alpha  \beta }} \right.
 \kern-\nulldelimiterspace} \beta }$, the probability of a channel being unavailable $\varepsilon $ and the probability of a channel being used $p$ are mutually dependent and correlated with each other in (\ref{8}), (\ref{9}), (\ref{10}) and (\ref{11}). Fortunately, above probability values can be obtained by solving linear equations if we know the outage probability ${p_{out}}\left( {{\gamma _0},\Delta } \right)$. In the next paragraph, we compute the value of the outage probability.

 \subsection{Outage Probability Model}
 Considering that in the typical cell ${{\cal C}_{ori}}$, the MU located at $x_i$ is associated with the BS located at $y_j$ and co-channel interference from adjacent BSs exists. Based on the general wireless propagation environments described in Section~\ref{sec2}, the SINR at the MU located at $x_i$ is expressed as follows
\[SIN{R_{{y_j}}}({x_i}) = \frac{{{K'} \cdot {S_{{y_j}}}({x_i}){{\left\{ {L\left( {\left\| {{y_j} - {x_i}} \right\|} \right)} \right\}}^{ - 1}}}}{{{\sigma ^2} + I_{{x_i}}^\Delta }},\tag{12a}\]
 with
 \begin{equation}\label{12b}
I_{{x_i}}^\Delta  = \sum\limits_{\scriptstyle{y_l} \in {\Theta _B}\backslash \{ {y_j}\} ,\hfill\atop
\scriptstyle\# \left\{ {{\Theta _B}\backslash \{ {y_j}\} } \right\} = \Delta \hfill} {{K'} \cdot {S_{{y_l}}}({x_i}){{\left\{ {L\left( {\left\| {{y_l} - {x_i}} \right\|} \right)} \right\}}^{ - 1}}},\tag{12b}
\end{equation}
where $I_{{x_i}}^\Delta $ is the aggregate interference received at the MU conditioned on that there are ${\Delta}$ interferers in total;
$\#\left\{ \cdot \right\}$ is an operation which counts the number of elements in a set; ${\sigma ^2}$ is the Gaussian noise power. Assume that all BSs transmit with the same power and as a result, ${K'} = K{P_{y_j}}$, ${y_j} \in {\Phi _B}$.

Because the sum of the outage probability ${p_{out}}\left( {{\gamma _0},\Delta } \right)$ and the success probability ${p_{suc}}\left( {{\gamma _0},\Delta } \right)$ equals to 1, ${p_{out}}\left( {{\gamma _0},\Delta } \right)$ is expressed by
\begin{equation}\label{eq13}
{p_{out}}\left( {{\gamma _0},\Delta } \right) = 1 - {p_{suc}}\left( {{\gamma _0},\Delta } \right){\kern 1pt}.\tag{13}
\end{equation}

When the MU is assumed to be associated with the nearest BS, the success probability of a MU located at ${x_i} \in {{\cal C}_{ori}}$ is derived as (\ref{eq14}),
\begin{figure*}[!t]
\[ \begin{array}{l}\label{eq14}
\begin{aligned}
{p_{suc}}\left( {{\gamma _0},\Delta } \right){\kern 1pt} {\kern 1pt} & = {\rm \mathbf{P}}\left( {\left. {SIN{R_{{y_j}}}\left( {{x_i}} \right) > {\gamma _0}} \right|{x_i} \in {{\cal C}_{ori}}} \right)\\
  &= {\rm \mathbf{P}}\left( {\left. {\frac{{{K'} \cdot {S_{{y_j}}}({x_i}){{\left\{ {L\left( {\left\| {{y_j} - {x_i}} \right\|} \right)} \right\}}^{ - 1}}}}{{{\sigma ^2} + {I_{{x_i},\Delta }}}} > {\gamma _0}} \right|{x_i} \in {{\cal C}_{ori}}} \right)\\
  &= {\rm \mathbf{P}}\left( {\left. {{S_{{y_j}}}({x_i}) > \frac{{{\gamma _0}}}{{{K'}}} \cdot L\left( {\left\| {{y_j} - {x_i}} \right\|} \right) \cdot \left( {{\sigma ^2} + I_{{x_i}}^\Delta } \right)} \right|{x_i} \in {{\cal C}_{ori}}} \right)\\
  &= {\rm \mathbf{E}}\left\{ {{\rm \mathbf{P}}\left( {\left. {{\sigma ^2} + I_{{x_i}}^\Delta  < \frac{{{K'}}}{{{\gamma _0}}} \cdot {S_{{y_j}}}({x_i}){{\left\{ {L\left( r \right)} \right\}}^{ - 1}}} \right|r} \right)} \right\}\\
  &= {\rm \mathbf{E}}\left\{ {\int_{r > 0} {2\pi {\lambda _B}r{e^{ - \pi {\lambda _B}{r^2}}}dr} \int_{ - \infty }^{ + \infty } {g\left( t \right) \cdot \mathbf{1}\left\{ {0 < t < \frac{{{K'}}}{{{\gamma _0}}} \cdot {S_{{y_j}}}({x_i}){{\left\{ {L\left( r \right)} \right\}}^{ - 1}}} \right\} \cdot dt} } \right\}\end{aligned}. \tag{14}
\end{array}\]
\end{figure*}
where ${\mathop{\rm \mathbf{E}}\nolimits} \left\{ \cdot \right\}$ is an expectation operation, $g\left( t \right)$ is the probability distribution function (PDF) of $\left( {{\sigma ^2} + I_{{x_i}}^\Delta } \right)$ and $\mathbf{1}\left\{ \cdot \right\}$ is an indicator function, which equals to 1 when the condition inside the bracket is satisfied and 0 otherwise. Moreover, the success probability can be further derived by using Parseval theorem \cite{Baccelli97} given below.

\mbox{\boldmath $Parseval$} \mbox{\boldmath $theorem$}: Denote the Laplace transforms of two square-integrable functions $f\left( t \right)$ and $g\left( t \right)$ as $F\left( \omega \right)$ and $G\left( \omega \right)$, respectively, i.e., $f\left( t \right) = \frac{1}{{2\pi }}\int_{ - \infty }^{ + \infty } {F\left( \omega  \right){e^{i\omega t}}d\omega } $ and $g\left( t \right) = \frac{1}{{2\pi }}\int_{ - \infty }^{ + \infty } {G\left( \omega  \right){e^{i\omega t}}d\omega } $, then they have the relationship as follows
\begin{equation}\label{15}
\int_{ - \infty }^{ + \infty } {f\left( t \right)g\left( t \right)\cdot dt}  = \frac{1}{{2\pi }}\int_{ - \infty }^{ + \infty } {F\left( \omega  \right)G\left( \omega  \right)\cdot d\omega }.\tag{15}
\end{equation}

Let the PDF of ${S_{{y_i}}}({x_i})$ be $f\left( y \right)$ and assume that $f\left( y \right)$ is independent of $g\left( y \right)$. Based on (\ref{15}) and the assumption that the noise and interference are independent of each other, (\ref{eq14}) can be further derived as (\ref{eq16}),
\begin{figure*}[!t]
\begin{equation}\label{eq16}
\begin{gathered}
  \begin{aligned}
  & \qquad {p_{suc}}\left( {{\gamma _0},\Delta } \right){\kern 1pt} {\kern 1pt} {\kern 1pt} {\kern 1pt}   \\
  & = {\rm \mathbf{E}}\left\{ {\int_{r > 0} {2\pi {\lambda _B}r{e^{ - \pi {\lambda _B}{r^2}}}dr}  \cdot \frac{1}{{2\pi }} \cdot \int_{ - \infty }^{ + \infty } {{\mathcal{L}_{{\sigma ^2}}}\left( { - 2\pi j\omega } \right) \cdot {\mathcal{L}_{I_{{x_i}}^\Delta }}\left( { - 2\pi j\omega } \right) \cdot \frac{{{e^{2\pi j\omega \frac{{K'}}{{{\gamma _0}}} \cdot {S_{{y_j}}}({x_i}){{\left\{ {L\left( r \right)} \right\}}^{ - 1}}}} - 1}}{{2\pi j\omega }}d\omega } } \right\} \hfill \\
  &= \int_{r > 0} {2\pi {\lambda _B}r{e^{ - \pi {\lambda _B}{r^2}}}dr} \int_{ - \infty }^{ + \infty } {f\left( y \right) \cdot dy} \int_{ - \infty }^{ + \infty } {{\mathcal{L}_{{\sigma ^2}}}\left( { - 2\pi j\omega } \right) \cdot {\mathcal{L}_{I_{{x_i}}^\Delta }}\left( { - 2\pi j\omega } \right) \cdot \frac{{{e^{2\pi j\omega \frac{{K'}}{{{\gamma _0}}} \cdot y \cdot {{\left\{ {L\left( r \right)} \right\}}^{ - 1}}}} - 1}}{{4{\pi ^2}j\omega }}d\omega }  \hfill \\
\end{aligned} \end{gathered}. \tag{16}
\end{equation}
\end{figure*}
where ${\mathcal{L}_{{\sigma ^2}}}\left( \cdot \right)$ and ${\mathcal{L}_{I_{{x_i}}^\Delta }}\left( \cdot \right)$ are the Laplace transforms of noise and interference $I_{{x_i}}^\Delta $, respectively. Assume that ${\int_0^{ \infty } {\int_{ - \infty }^{ \infty } {\left| {{{\frac{f\left( y \right)}{4{\pi ^2}j\omega } \left\{ {\exp \left[ {2\pi j\omega y \frac{{K'}}{{{\gamma _0}}}  {{\left\{ {L\left( r \right)} \right\}}^{ - 1}}} \right] - 1} \right\}}  }} \right|d\omega dy} }}$
 is finite, then the success probability can be obtained as (\ref{eq17}). Further with the assumption of (\ref{eq18}), (\ref{eq17}) can be further derived as (\ref{eq19}), where ${\mathcal{L}_{{S_{{y_i}}}({x_i})}}\left( \cdot \right)$ is the Laplace transform of ${S_{{y_j}}}({x_i})$. Let $s =  - \omega K' \cdot {\left\{ {{\gamma _0}L\left( r \right)} \right\}^{ - 1}}$ and (\ref{eq19}) is expressed by (\ref{eq20}).

  \begin{figure*}[!t]
 \begin{equation}\label{eq17}
\begin{gathered}\begin{aligned}
  & \qquad {p_{suc}}\left( {{\gamma _0},\Delta } \right){\kern 1pt} {\kern 1pt} {\kern 1pt} {\kern 1pt}  \\
  & = \int_{r > 0} {2\pi {\lambda _B}r{e^{ - \pi {\lambda _B}{r^2}}}dr} \int_{ - \infty }^{ + \infty } {f\left( y \right) \cdot \frac{{{e^{2\pi j\omega \frac{{K'}}{{{\gamma _0}}} \cdot y \cdot {{\left\{ {L\left( r \right)} \right\}}^{ - 1}}}} - 1}}{{4{\pi ^2}j\omega }} \cdot dy} \int_{ - \infty }^{ + \infty } {{\mathcal{L}_{{\sigma ^2}}}\left( { - 2\pi j\omega } \right) \cdot {\mathcal{L}_{I_{{x_i}}^\Delta }}\left( { - 2\pi j\omega } \right)d\omega }  \hfill \\
  &= \int_{r > 0} {2\pi {\lambda _B}r{e^{ - \pi {\lambda _B}{r^2}}}dr}  \cdot  \hfill
  \int_{ - \infty }^{ + \infty } {\frac{{\int_{ - \infty }^{ + \infty } {f\left( y \right) \cdot {e^{2\pi j\omega \frac{{K'}}{{{\gamma _0}}} \cdot y \cdot {{\left\{ {L\left( r \right)} \right\}}^{ - 1}}}}dy}  - \int_{ - \infty }^{ + \infty } {f\left( y \right)dy} }}{{4{\pi ^2}j\omega }}d\omega } \\ & \qquad \times \int_{ - \infty }^{ + \infty } {{\mathcal{L}_{{\sigma ^2}}}\left( { - 2\pi j\omega } \right) \cdot {\mathcal{L}_{I_{{x_i}}^\Delta }}\left( { - 2\pi j\omega } \right)d\omega }  \hfill \\
  \end{aligned}\end{gathered}.\tag{17}
  \end{equation}
  \end{figure*}

\begin{figure*}[!t]
\begin{equation}\label{eq18}
\int_{r > 0} {\int_{ - \infty }^{ + \infty } {\left| {\frac{{\int_{ - \infty }^{ + \infty } {f\left( y \right) \cdot {e^{2\pi j\omega \frac{{K'}}{{{\gamma _0}}} \cdot y \cdot {{\left\{ {L\left( r \right)} \right\}}^{ - 1}}}}dy}  - \int_{ - \infty }^{ + \infty } {f\left( y \right)dy} }}{{4{\pi ^2}j\omega }}} \right| \cdot \left| {{\mathcal{L}_{{\sigma ^2}}}\left( { - 2\pi j\omega } \right)} \right| \cdot \left| {{\mathcal{L}_{I_{{x_i}}^\Delta }}\left( { - 2\pi j\omega } \right)} \right|d\omega } dr}  < \infty .\tag{18}
\end{equation}
\end{figure*}

\begin{figure*}
\begin{equation}\label{eq19}
\begin{gathered}\begin{aligned}
  & \qquad {p_{suc}}\left( {{\gamma _0},\Delta } \right){\kern 1pt} {\kern 1pt} {\kern 1pt} {\kern 1pt}  \\
  &= {\kern 1pt} \int_{r > 0} {2\pi {\lambda _B}r{e^{ - \pi {\lambda _B}{r^2}}}dr}  \cdot  \hfill
 \int_{ - \infty }^{ + \infty } {\frac{{\int_{ - \infty }^{ + \infty } {f\left( y \right) \cdot {e^{2\pi j\omega \frac{{K'}}{{{\gamma _0}}} \cdot y \cdot {{\left\{ {L\left( r \right)} \right\}}^{ - 1}}}}dy}  - \int_{ - \infty }^{ + \infty } {f\left( y \right)dy} }}{{4{\pi ^2}j\omega }} \cdot {\mathcal{L}_{{\sigma ^2}}}\left( { - 2\pi j\omega } \right) \cdot {\mathcal{L}_{I_{{x_i}}^\Delta }}\left( { - 2\pi j\omega } \right)d\omega }  \hfill \\
  & = \int_{r > 0} {2\pi {\lambda _B}r{e^{ - \pi {\lambda _B}{r^2}}}}
   \int_{ - \infty }^{ + \infty } {\frac{{{\mathcal{L}_{{S_{{y_j}}}({x_i})}}\left( { - 2\pi j\omega K' \cdot {{\left\{ {{\gamma _0}L\left( r \right)} \right\}}^{ - 1}}} \right) - {\mathcal{L}_{{S_{{y_j}}}({x_i})}}\left( 0 \right)}}{{4{\pi ^2}j\omega }} \cdot {\mathcal{L}_{{\sigma ^2}}}\left( { - 2\pi j\omega } \right) \cdot {\mathcal{L}_{I_{{x_i}}^\Delta }}\left( { - 2\pi j\omega } \right)d\omega dr}  \hfill \\
  \end{aligned} \end{gathered}. \tag{19}
  \end{equation}
 \end{figure*}

\begin{figure*}
\begin{equation}\label{eq20}
\begin{aligned}& \qquad {p_{suc}}\left( {{\gamma _0},\Delta } \right)  \\
&= \int_{r > 0} {2\pi {\lambda _B}r{e^{ - \pi {\lambda _B}{r^2}}}}
  \int_{ - \infty }^{ + \infty } {\frac{{{\mathcal{L}_{{S_{{y_j}}}({x_i})}}\left( {2\pi js} \right) - {\mathcal{L}_{{S_{{y_j}}}({x_i})}}\left( 0 \right)}}{{4{\pi ^2}js}} \cdot {\mathcal{L}_{{\sigma ^2}}}\left( {{{2\pi j{\gamma _0}L\left( r \right)s} \mathord{\left/
 {\vphantom {{2\pi j{\gamma _0}L\left( r \right)s} {K'}}} \right.
 \kern-\nulldelimiterspace} {K'}}} \right) \cdot {\mathcal{L}_{I_{{x_i}}^\Delta }}\left( {{{2\pi j{\gamma _0}L\left( r \right)s} \mathord{\left/
 {\vphantom {{2\pi j{\gamma _0}L\left( r \right)s} {K'}}} \right.
 \kern-\nulldelimiterspace} {K'}}} \right)ds} dr \end{aligned}.\tag{20}
 \end{equation}
 \end{figure*}

The Laplace transform of the noise is expressed as \cite{Baccelli97}
 \[{\mathcal{L}_{{\sigma ^2}}}\left( s \right) = {e^{ - {\sigma ^2}s}}.\tag{21}\]

 According to the definition of $I_{{x_i}}^\Delta $ in (\ref{12b}), we follow a basic technique in \cite{Sousa90} to obtain the outage probability condition on having $\Delta$ interferers, which consists of two steps:
 \begin{enumerate}
 \item Consider a finite network, say on disk of radius $a$ centered at the origin, and condition on having a constant number of nodes in this finite area, for example ${\Delta}$ nodes. Assume that the nodes¡¯ locations are independent identical distribution.
 \item Let the disk radius go to infinity, while keeping the node density, i.e. the ratio of the number of nodes to the network area, constant.
 \end{enumerate}

$\mathbf{Step 1}$: Conditioning on having ${\Delta}$ nodes (i.e., interferers) in the disk of radius $a$, the Laplace transform of aggregate interference is denoted by
\[{\mathcal{L}_{I_{{x_i}}^\Delta ,a}}\left( s \right) \triangleq {\rm \mathbf{E}}\left\{ {\left. {{e^{ - s{I_{{x_i},\Delta ,a}}}}} \right|\# B\left( {ori,a} \right) = \Delta } \right\},\tag{22}\]
where $B\left( {ori,a} \right)$ is the set of nodes located on a disk with radius $a$ centered at the origin $ori$ and $\# B\left( {ori,a} \right) = \Delta $ means that the number of nodes in $B\left( {ori,a} \right)$ is $\Delta$. These nodes (i.e., interferers) are uniformly distributed on the disk with radial density
\[{f_R}\left( r \right) = \left\{ \begin{gathered}
  \frac{{2r}}{{{a^2}}}{\kern 1pt} {\kern 1pt} {\kern 1pt} {\kern 1pt} {\kern 1pt} {\kern 1pt} {\kern 1pt} if{\kern 1pt} {\kern 1pt} 0 \leqslant r \leqslant a \hfill \\
  0{\kern 1pt} {\kern 1pt} {\kern 1pt} {\kern 1pt} {\kern 1pt} {\kern 1pt} {\kern 1pt} {\kern 1pt} {\kern 1pt} {\kern 1pt} {\kern 1pt} {\kern 1pt} {\kern 1pt} {\kern 1pt} otherwise \hfill \\
\end{gathered}  \right..\tag{23}\]

Therefore, the Laplace transform ${\mathcal{L}_{I_{{x_i}}^\Delta ,a}}\left( s \right)$ is the product of the $\Delta$ individual Laplace transform, which is given by (\ref{eq24}).
\begin{figure*}[!t]
\begin{equation}\label{eq24}
\begin{gathered}
  {\mathcal{L}_{I_{{x_i}}^\Delta ,a}}\left( s \right){\text{ = }}{\rm \mathbf{E}}\left\{ {\left. {{e^{ - s{I_{{x_i},\Delta ,a}}}}} \right|\# B\left( {ori,a} \right) = \Delta } \right\} \hfill = {\rm \mathbf{E}}\left\{ {{{\left( {\int_0^a {\frac{{2r}}{{{a^2}}}\exp \left( { - sK' \cdot {S_{{y_l}}}({x_i}){{\left\{ {L\left( r \right)} \right\}}^{ - 1}}} \right)dr} } \right)}^\Delta }} \right\}{\kern 1pt} {\kern 1pt} {\kern 1pt} {\kern 1pt} {\kern 1pt} {\kern 1pt}  \hfill \\
\end{gathered}. \tag{24}
\end{equation}
\end{figure*}

Assume that the path loss law is $L\left( r \right){\text{ = }}{r^b}$, then the individual Laplace transform is derived as follows \cite{Jeffrey07}
\begin{equation}\label{25}
\begin{gathered}
  {\kern 1pt} {\kern 1pt} {\kern 1pt} {\kern 1pt} {\kern 1pt} {\kern 1pt} {\kern 1pt} {\kern 1pt} {\kern 1pt} {\kern 1pt} {\kern 1pt} {\kern 1pt} {\kern 1pt} \int_0^a {\frac{{2r}}{{{a^2}}}\exp \left( { - sK' \cdot {S_{{y_l}}}({x_i}){{\left\{ {L\left( r \right)} \right\}}^{ - 1}}} \right)dr}  \hfill \\
   = \int_0^a {\frac{{2r}}{{{a^2}}}\exp \left( { - sK' \cdot {S_{{y_l}}}({x_i}){r^{ - b}}} \right)dr}  \hfill \\
   =  - \frac{2}{{{a^2}}}\int_{\frac{1}{a}}^\infty  {{y^{ - 3}}\exp \left( { - sK' \cdot {S_{{y_l}}}({x_i}){y^b}} \right)dy}  \hfill \\
   = \frac{1}{b}{\left[ {s{K'} \cdot {S_{{y_l}}}({x_i})} \right]^{\frac{2}{b}}} \cdot \Gamma \left( { - \frac{2}{b},s{K'}\frac{1}{{{a^b}}}{S_{{y_l}}}({x_i})} \right) \hfill \\
\end{gathered}, \tag{25}
\end{equation}
where $\Gamma \left( {\cdot} \right)$ denotes Gamma function. \\

$\mathbf{Step 2}$: When $a$ goes to infinity, (\ref{25}) is further derived by
\begin{equation}
\begin{gathered}
  {\kern 1pt} {\kern 1pt} {\kern 1pt} {\kern 1pt} {\kern 1pt} {\kern 1pt} {\kern 1pt} {\kern 1pt} {\kern 1pt} {\kern 1pt} {\kern 1pt} {\kern 1pt} {\kern 1pt} \mathop {\lim }\limits_{a \to \infty } \int_0^a {\frac{{2r}}{{{a^2}}}\exp \left( { - sK' \cdot {S_{{y_l}}}({x_i}){{\left\{ {L\left( r \right)} \right\}}^{ - 1}}} \right)dr}  \hfill \\
   = \frac{1}{b}{\left[ {sK' \cdot {S_{{y_l}}}({x_i})} \right]^{\frac{2}{b}}} \cdot \Gamma \left( { - \frac{2}{b},0} \right) \hfill \\
   = \frac{1}{b}{\left[ {sK' \cdot {S_{{y_l}}}({x_i})} \right]^{\frac{2}{b}}} \cdot \Gamma \left( { - \frac{2}{b}} \right){\kern 1pt} {\kern 1pt} {\kern 1pt} {\kern 1pt}  \hfill \\
\end{gathered}. \tag{26}
\end{equation}

Then ${\mathcal{L}_{{I_{{x_i},\Delta }}}}\left( s \right)$ is  is transformed into (\ref{eq27}). By substituting (\ref{eq20}) and (\ref{eq27}) into (\ref{eq13}), the outage probability conditioned on having $\Delta$ interferers is derived by (\ref{eq28}).
\begin{figure*}[!t]
\begin{equation}\label{eq27}
\begin{gathered}
\begin{aligned}
  {\mathcal{L}_{I_{{x_i}}^\Delta ,a}}\left( s \right) &= \mathop {\lim }\limits_{a \to \infty } {\mathcal{L}_{{I_{{x_i},\Delta ,a}}}}\left( s \right){\text{ = }}{{\rm \mathbf{E}}_{{S_{{y_l}}}({x_i})}}{\left( {\frac{1}{b}{{\left[ {sK' \cdot {S_{{y_l}}}({x_i})} \right]}^{\frac{2}{b}}} \cdot \Gamma \left( { - \frac{2}{b}} \right)} \right)^\Delta }  \\
  &= {\left( {\frac{1}{b}{{\left( {sK'} \right)}^{\frac{2}{b}}} \cdot \Gamma \left( { - \frac{2}{b}} \right)} \right)^\Delta }{{\rm \mathbf{E}}_{{S_{{y_l}}}({x_i})}}\left[ {{{\left( {{S_{{y_l}}}({x_i})} \right)}^{\frac{{2\Delta }}{b}}}} \right]
  \end{aligned}
\end{gathered}. \tag{27}
\end{equation}
\end{figure*}

\begin{figure*}[!t]
\begin{equation}\label{eq28}
\begin{gathered}\begin{aligned}
&\quad {p_{out}}\left( {{\gamma _0},\Delta } \right) \\
&= 1 - \int_{r > 0} {2\pi {\lambda _B}r{e^{ - \pi {\lambda _B}{r^2}}}}
  {\kern 1pt} \int_{ - \infty }^{ + \infty } {\frac{{{\mathcal{L}_{{S_{{y_j}}}({x_i})}}\left( {2\pi js} \right) - {\mathcal{L}_{{S_{{y_j}}}({x_i})}}\left( 0 \right)}}{{4{\pi ^2}js}} \cdot {\mathcal{L}_{{\sigma ^2}}}\left( {{{2\pi j{\gamma _0}L\left( r \right)s} \mathord{\left/
 {\vphantom {{2\pi j{\gamma _0}L\left( r \right)s} {K'}}} \right.
 \kern-\nulldelimiterspace} {K'}}} \right)} \\
 & \quad \times {\mathcal{L}_{I_{{x_i}}^\Delta }}\left( {{{2\pi j{\gamma _0}L\left( r \right)s} \mathord{\left/
 {\vphantom {{2\pi j{\gamma _0}L\left( r \right)s} {K'}}} \right.
 \kern-\nulldelimiterspace} {K'}}} \right)ds dr  \\
  &= 1 -  \int_{r > 0} {2\pi {\lambda _B}r{e^{ - \pi {\lambda _B}{r^2}}}}
  \int_{ - \infty }^{ + \infty } {\frac{{{\mathcal{L}_{{S_{{y_j}}}({x_i})}}\left( {2\pi js} \right) - {\mathcal{L}_{{S_{{y_j}}}({x_i})}}\left( 0 \right)}}{{4{\pi ^2}js}} \cdot \exp \left( { - \frac{{2\pi j{\gamma _0}{\sigma ^2}{r^{ - b}}s}}{{K'}}} \right)}  \\
  & \quad \times {\left( {\frac{1}{b}{{\left( {2\pi j{\gamma _0}{r^{ - b}}s} \right)}^{\frac{2}{b}}} \cdot \Gamma \left( { - \frac{2}{b}} \right)} \right)^\Delta }{{\rm \mathbf{E}}_{{S_{{y_l}}}({x_i})}}\left[ {{{\left( {{S_{{y_l}}}({x_i})} \right)}^{\frac{{2\Delta }}{b}}}} \right]dsdr   \\
\end{aligned}\end{gathered} .\tag{28}
\end{equation}
\end{figure*}

\subsection{Blocking Probability with Rayleigh Fading}
When wireless signals are assumed to suffer Rayleigh fadings, the PDF of signal is expressed as
\[f(x) = x/{\sigma ^2} \cdot \exp \left( { - {x^2}/2{\sigma ^2}} \right){\kern 1pt} {\kern 1pt} (0 \leqslant x \leqslant \infty ).\tag{29}\]
And the PDF of the signal power is expressed as
\[f(y) = 1/2{\sigma ^2} \cdot \exp \left( { - y/2{\sigma ^2}} \right){\kern 1pt} {\kern 1pt} (0 \leqslant y \leqslant \infty ).\tag{30}\]
It is obvious that the PDF of the signal power follows an exponential distribution. Without loss of generality, let ${S_{{y_l}}}({x_i}) \sim \exp \left( 1 \right)$. Then the Laplace transform of ${S_{{y_l}}}({x_i})$ is derived by
\[{\mathcal{L}_{{S_{{y_l}}}({x_i})}}\left( s \right) = \frac{1}{{s + 1}},\tag{31}\]
and then
\[{\rm \mathbf{E}}\left\{ {{{\left( {{S_{{y_l}}}({x_i})} \right)}^{\frac{{2\Delta }}{b}}}} \right\} = \Gamma \left( {1 + \frac{{2\Delta }}{b}} \right).\tag{32}\]
Moreover, the Laplace transform of the aggregate interference is derived by
\begin{equation}\label{33}
{\mathcal{L}_{I_{{x_i}}^\Delta ,a}}\left( s \right) = {\left( {\frac{1}{b}{{\left( {sK'} \right)}^{\frac{2}{b}}} \cdot \Gamma \left( { - \frac{2}{b}} \right)} \right)^\Delta } \cdot \Gamma \left( {1 + \frac{{2\Delta }}{b}} \right).\tag{33}
\end{equation}
By substituting (\ref{33}) into (\ref{eq28}), the outage probability is derived as (\ref{eq34}).
\begin{figure*}[!t]
\begin{equation}\label{eq34}
\begin{gathered}\begin{aligned} \quad {p_{out}}\left( {{\gamma _0},\Delta } \right){\kern 1pt} {\kern 1pt} {\kern 1pt} {\kern 1pt}  &= 1 - \int_{r > 0} {\int_{ - \infty }^{ + \infty } { - \frac{{{\lambda _B}r{e^{ - \pi {\lambda _B}{r^2}}}}}{{2\pi js + 1}}}  \cdot \exp \left( { - {{2\pi j{\gamma _0}{\sigma ^2}{r^{ - b}}s} \mathord{\left/
 {\vphantom {{2\pi j{\gamma _0}{\sigma ^2}{r^{ - b}}s} {K'}}} \right.
 \kern-\nulldelimiterspace} {K'}}} \right)  }  \\
  & \quad \times{\left( {\frac{1}{b}{{\left( {2\pi {\gamma _0}{r^{ - b}}js} \right)}^{\frac{2}{b}}} \cdot \Gamma \left( { - \frac{2}{b}} \right)} \right)^\Delta }\Gamma \left( {1 + \frac{{2\Delta }}{b}} \right)dsdr  \\
\end{aligned}\end{gathered}. \tag{34}
\end{equation}
\end{figure*}

Assume that a call will be dropped if the call can not be served immediately in PVT random cellular networks. As a consequence, if the total number of active MUs exceeds the number of available channels in the typical cell ${\mathcal{C}_{ori}}$, a call will be blocked due to a lack of sufficient channel resource. Therefore, the blocking probability is derived by
 \[{p_b} = \sum\limits_{m = n \leqslant C} {\pi (m,n)}  = \sum\limits_{i = j \leqslant C} {\frac{1}{\chi }{{\left( {\frac{\lambda }{\eta }} \right)}^m}\frac{1}{{m!}}{\binom{C}{n}}{{\left( {\frac{\alpha }{\beta }} \right)}^n}},\tag{35a} \]
with
\[\frac{\alpha }{\beta }{\text{ = }}\frac{{1 - \varepsilon }}{\varepsilon },\tag{35b}\]
\[\varepsilon  = \mathop {\lim }\limits_{{N_I} \to \infty } \sum\limits_{\vartriangle  = 1}^{{N_I}} {{p_{out}}\left( {{\gamma _0},\Delta } \right){\binom{N_I}{\Delta}}{p^\Delta }{{\left( {1 - p} \right)}^{{N_I} - \Delta }}},\tag{35c} \]
\begin{figure*}
\begin{equation}
\begin{gathered}\begin{aligned} \quad {p_{out}}\left( {{\gamma _0},\Delta } \right){\kern 1pt} {\kern 1pt} {\kern 1pt} {\kern 1pt}  &= 1 - \int_{r > 0} {\int_{ - \infty }^{ + \infty } { - \frac{{{\lambda _B}r{e^{ - \pi {\lambda _B}{r^2}}}}}{{2\pi js + 1}}}  \cdot \exp \left( { - {{2\pi j{\gamma _0}{\sigma ^2}{r^{ - b}}s} \mathord{\left/
 {\vphantom {{2\pi j{\gamma _0}{\sigma ^2}{r^{ - b}}s} {K'}}} \right.
 \kern-\nulldelimiterspace} {K'}}} \right)  }  \\
  & \quad \times{\left( {\frac{1}{b}{{\left( {2\pi {\gamma _0}{r^{ - b}}js} \right)}^{\frac{2}{b}}} \cdot \Gamma \left( { - \frac{2}{b}} \right)} \right)^\Delta }\Gamma \left( {1 + \frac{{2\Delta }}{b}} \right)dsdr  \\
\end{aligned}\end{gathered}. \tag{35d}
\end{equation}
\end{figure*}

Furthermore, the mean sojourn time of a MU can be derived by Little's theorem \cite{Cooper81}
 \[D = \frac{N}{\lambda },\tag{36}\]
 where $N = \sum\limits_{m = 1}^C {m \cdot \pi \left( {m,n} \right)} $ is the mean number of served MUs in the typical cell ${\mathcal{C}_{ori}}$.

 \subsection{Performance Analysis}
 Assuming Rayleigh fading channels, the blocking probability and the mean sojourn time of PVT random cellular networks can be numerically computed. Following the simulation configuration in \cite{Anand03,Dhillon12} and \cite{Xiang13}, default parameters for a PVT random cellular network are configured as follows: BS intensity is ${\lambda _B} = 0.2$ per square kilometers; the maximum number of available channels in a typical cell ${\mathcal{C}_{ori}}$ is $C = 20$; the arrival rate of calls is $\lambda  = 1 {minute^{-1}}$; the BS transmitting power is ${P_{{y_i}}} = 30$dBm; the mean noise power is ${\sigma ^2} = 0$dBm; the path loss exponent is $b = 4$; the antenna gain is $K = 31.54$dB for an urban microcell environment.

\begin{figure}
\centerline{\includegraphics[width=7.5cm,draft=false]{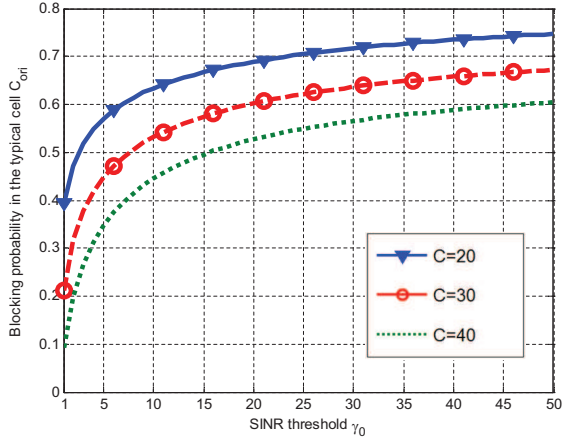}}
\caption{\small Blocking probability versus SINR threshold with different maximum channel numbers $C$.}
\label{fig3}
\end{figure}

\begin{figure}
\centerline{\includegraphics[width=7.5cm,draft=false]{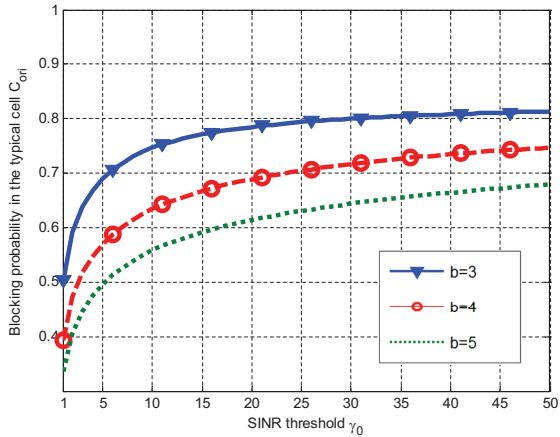}}
\caption{\small Blocking probability versus SINR threshold with different $b$.}
\label{fig4}
\end{figure}

\begin{figure}
\centerline{\includegraphics[width=7.5cm,draft=false]{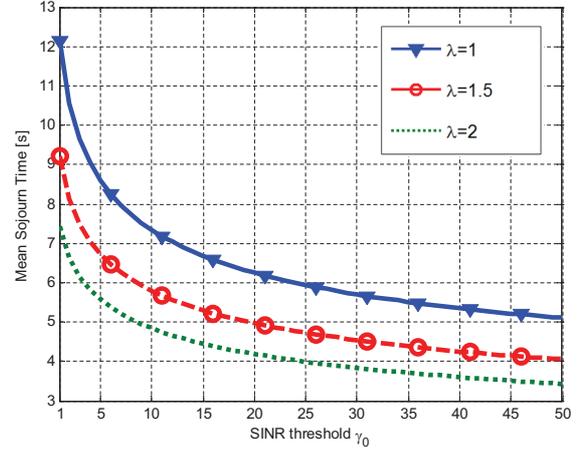}}
\caption{\small Mean sojourn time versus SINR threshold with different $\lambda$.}
\label{fig5}
\end{figure}

Fig.~\ref{fig3} shows the call blocking probability with respect to the SINR threshold considering different maximum number of channel numbers $C$ in a typical cell ${\mathcal{C}_{ori}}$. When the maximum number of available channels is fixed, it is observed that the blocking probability increases with the increase of SINR threshold. The reason is that the successful decoded signal is reduced for MUs when the SINR threshold is increased. When the SINR threshold is fixed, the blocking probability increases with the decrease of the maximum available channels number in the cell ${\mathcal{C}_{ori}}$. That is because a call is more likely to be dropped when channel resource is insufficient.

Fig.~\ref{fig4} illustrates the blocking probability versus the SINR threshold for different path loss exponents $b$ in the typical cell ${\mathcal{C}_{ori}}$. When the SINR threshold is fixed, the blocking probability decreases with the increase of the path loss exponent. It is well known that the path loss exponent affects both desired signals and interference signals. However, these curves imply that the path loss exponent has a more significant attenuation on the aggregated interference in PVT random cellular networks.

Fig.~\ref{fig5} shows the mean sojourn time versus the SINR threshold for different arrival rates of calls. When the SINR threshold is fixed, the mean sojourn time decreases with the increase of arrival rate of calls. When the arrival rate of calls are fixed, the mean sojourn time decreases with the increase of SINR threshold. That is because the number of available channels decreases with the increase of SINR threshold. As a result, the service ability of cellular network is decreased and then the mean sojourn time of a MU ¡°staying¡± in the PVT random cellular network is decreased.

\section{Spatial Spectrum and Energy Efficiency of PVT Random Cellular Networks}
\label{sec4}
In this section, we further evaluate the spatial spectrum and energy efficiency of PVT random cellular networks.

\subsection{Spatial Spectrum and Energy Efficiency}
Assume that the bandwidth of a typical cell ${\mathcal{C}_{ori}}$ is $B$, the throughput of the typical cell ${\mathcal{C}_{ori}}$ is then given by
\begin{figure*}[!t]
\begin{equation}\label{eq37}
{T_{throughput}} = \left( {1 - {p_b}} \right)B \cdot \operatorname{\rm \mathbf{E}} \left\{ {{{\log }_2}\left( {1 + SIN{R_{{y_j}}}\left( {{x_i}} \right)} \right)} \right\} \cdot \sum\limits_{0 \leqslant m \leqslant n \leqslant C} {m \cdot \pi (m,n)}.\tag{37}
\end{equation}
\end{figure*}
Without loss of generality, the number of interferers on a wireless link of PVT random cellular networks is assumed as $\Delta$. Therefore, the link capacity between a MU and the associated BS is defined as
\begin{equation}\label{38}
\begin{gathered}
  \operatorname{\rm \mathbf{E}} \left\{ {{{\log }_2}\left( {1 + SIN{R_{{y_j}}}\left( {{x_i}} \right)} \right)} \right\} \hfill \\
   = \int_0^{ + \infty } {\rm \mathbf{P}\left( {{{\log }_2}\left( {1 + SIN{R_{{y_j}}}\left( {{x_i}} \right)} \right) > t} \right)} dt \hfill \\
   = \int_0^{ + \infty } {\rm \mathbf{P}\left( {SIN{R_{{y_j}}}\left( {{x_i}} \right) > {2^t} - 1} \right)} dt \hfill \\
   = \int_0^{ + \infty } {\left[ {1 - {P_{out}}\left( {{2^t} - 1,\Delta } \right)} \right]} dt\quad \hfill \\
\end{gathered}. \tag{38}
\end{equation}
A simple proof of (\ref{38}) is given as follows.

$\mathbf{Proof}$: Consider a random value $x$ which has continues PDF, then its expectation is given by
\begin{equation}\label{eq39}
\begin{aligned} {\rm \mathbf{E}}\left\{ x \right\} &= \int_{ - \infty }^{ + \infty } {xf(x)} dx \\
&= \int_{ - \infty }^0 {xf(x)} dx + \int_0^{ + \infty } {xf(x)} dx \end{aligned}.\tag{39}
\end{equation}
Furthermore, the two terms of the right side of (\ref{eq39}) are written as
\begin{equation}
\begin{aligned}& \quad \int_{ - \infty }^0 {xf(x)} dx =  - \int_{ - \infty }^0 {\left( {\int_x^0 {dy} } \right)f(x)} dx \\
&=  - \int_{ - \infty }^0 {\int_{ - \infty }^y {f(x)dxdy} }  =  - \int_{ - \infty }^0 {F(y)dy} \end{aligned}, \tag{40a}
\end{equation}
\begin{equation}\label{eq40b}
\begin{aligned}& \quad \int_0^{ + \infty } {xf(x)} dx = \int_0^{ + \infty } {\left( {\int_0^x {dy} } \right)f(x)} dx \\
&= \int_0^{ + \infty } {\int_y^{ + \infty } {f(x)dxdy} }  = \int_0^{ + \infty } {\left[ {1 - F(y)} \right]dy} \end{aligned}, \tag{40b}
\end{equation}
where $F\left(  \cdot  \right)$ denotes the cumulative probability function (CDF) operation. As we know
 $SIN{R_{{y_j}}}\left( {{x_i}} \right) \geqslant 0$. Substituting $SIN{R_{{y_j}}}\left( {{x_i}} \right)$ into (\ref{eq40b}), the result of (\ref{38}) is obtained.
Based on (\ref{eq37}), the spatial spectrum efficiency of the PVT random cellular network is derived as (\ref{eq41}) according to \cite{Soh13}, where ${\lambda_B}$ is the BS density of PVT random cellular networks.
\begin{figure*}[!t]
\begin{equation}\label{eq41}
\begin{gathered}
\begin{aligned}
  SSE &= {\lambda _B} \cdot {T_{throughput}}  \\
  &{\text{ = }}\left( {1 - {p_b}} \right)B{\lambda _B} \cdot \int_0^{ + \infty } {\left[ {1 - {P_{out}}\left( {{2^t} - 1,\Delta } \right)} \right]} dt \cdot \sum\limits_{0 \leqslant m \leqslant n \leqslant C} {m \cdot \pi (m,n)}
  \end{aligned}
\end{gathered},\tag{41}
\end{equation}
\end{figure*}

The energy efficiency of the typical cell ${\mathcal{C}_{ori}}$ during the whole life time is derived as follows
\[\varphi {\text{ = }}\frac{{{E_{total}}}}{{{D_{total}}}},\tag{42}\]
where $E_{total}$ denotes the BS energy consumption in the life time and $D_{total}$ denotes the BS throughput in the life time.
In the entire life time, the BS energy consumption includes the operation energy and the embodied energy \cite{Humar11}. Moreover, the embodied energy includes the initial embodied energy consumed in factories and the maintenance embodied energy in the life time. Therefore, the BS energy consumption in the life time is expressed as
\[{E_{total}} = {E_{EMinit}} + {E_{EMma\operatorname{int} }} + {E_{EMoper}},\tag{43}\]
where ${E_{EMinit}}$ denotes the initial embodied energy, ${E_{EMma\operatorname{int} }}$ denotes the maintenance embodied energy, and ${E_{EMoper}}$ is the operation energy, which is a linear function of the total transmission power over all occupied channels and is given as follows \cite{Richter09}
\begin{equation}
\begin{aligned}
{E_{EMoper}} &= \left[ {h \cdot {P_{chl}}\sum\limits_{0 \leqslant m \leqslant n \leqslant C} {m \cdot \pi \left( {m,n} \right)}  + k} \right]\\ &\times {t_{lifetime}}\quad ,
\end{aligned}\tag{44}
\end{equation}
where ${P_{chl}}$ is the transmission power over a wireless channel, ${t_{lifetime}}$ is the life time of a BS, $h$ and $k$ are linear coefficients of the total transmission power. The total throughput of BS in the life time is expressed as

\[{D_{total}} = {t_{lifetime}} \cdot {T_{throughput}}.\tag{45}\]

Therefore, the energy efficiency of the typical cell ${\mathcal{C}_{ori}}$ is derived as (\ref{eq46}).
\begin{figure*}[!t]
\begin{equation}\label{eq46}
\varphi  = \frac{{{t_{lifetime}} \cdot {T_{throughput}}}}{{{E_{EMinit}} + {E_{EMma\operatorname{int} }} + \left[ {h \cdot {P_{chl}}\sum\limits_{0 \leqslant m \leqslant n \leqslant C} {m \cdot \pi \left( {m,n} \right)}  + k} \right] \cdot {t_{lifetime}}}}.\tag{46}
\end{equation}
\end{figure*}
Based on the Palm theory, the result of (\ref{eq46}) can be extended to the whole PVT random cellular network.

\subsection{Numerical Results and Discussion}
Based on the spatial spectrum and energy efficiency analysis, numerical results are illustrated in this subsection. Based on default parameters used in Section~\ref{sec3}.D, some parameters used for comparing grid cellular networks (i.e., regular hexagonal cellular networks) and PVT random cellular networks are configured as follows \cite{Xiang13, Humar11, Richter09}: the channel bandwidth is $B = 0.1 $MHz, the embodied energy is configured as ${{E_{EMinit}} + {E_{EMma\operatorname{int} }}} = 85$GJ, the transmission power over a wireless channel is ${{P_{chl}} = 1 }$Watt, $h = 7.84$, $k = 71.5$, without loss of generality, the BS life time ${t_{lifetime}}$ is configured as 1 year, e.g., 365 days.

\begin{figure}
\centerline{\includegraphics[width=7.5cm,draft=false]{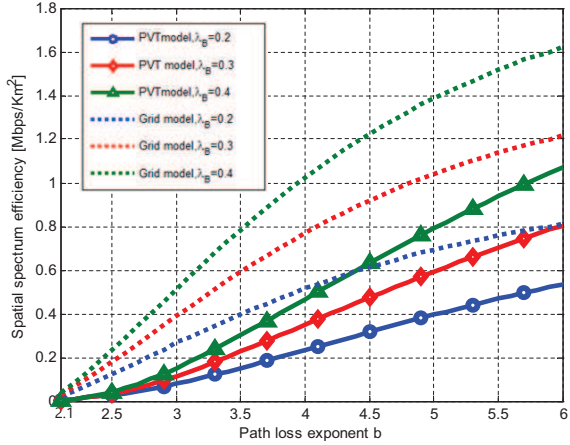}}
\caption{\small Spatial spectrum efficiency with respect to the path loss exponent $b$ considering different BS densities ${\lambda_B}$ in PVT and grid cellular networks.}
\label{fig6}
\end{figure}

Fig.~\ref{fig6} illustrates the spatial spectrum efficiency of PVT and grid cellular networks with respect to the path loss exponent and the BS density in cellular networks, in which ``PVT model" labels the PVT random cellular network and ``Grid model" represents the regular hexagonal cellular network. When the BS density ${\lambda_B}$ is fixed, numerical results of PVT and grid cellular networks consistently show that the spatial spectrum efficiency increases with the increase of the path loss exponent. When the path loss exponent $b$ is fixed, numerical results of PVT and grid cellular networks consistently illustrate that the spatial spectrum efficiency increases with the increase of the BS density. Compared with PVT model results and grid model results in Fig.~\ref{fig6}, it is shown that values corresponding to PVT random cellular networks are obviously less than values corresponding to grid cellular networks. This result is confirmed in \cite{Andrews11} which showed that the average transmission rate of PPP random cellular networks is less than the average transmission rate of grid cellular networks. Considering that the PVT random cellular network forms a special case of the PPP random cellular networks, the spatial spectrum efficiency of PVT random cellular network is less than the spatial spectrum efficiency of grid cellular network when the transmission bandwidth is fixed in PVT and grid cellular networks.

\begin{figure}
\centerline{\includegraphics[width=7.5cm,draft=false]{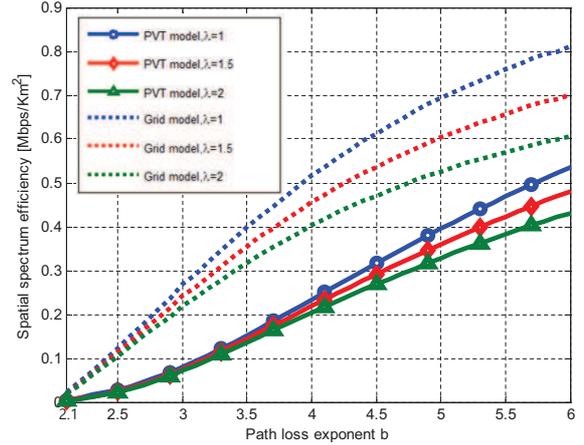}}
\caption{\small Spatial spectrum efficiency with respect to the path loss exponent $b$ considering different call arrival rates $\lambda$ in PVT and grid cellular networks.}
\label{fig7}
\end{figure}

In Fig.~\ref{fig7}, the effect of the call arrival rate $\lambda$ on the spatial spectrum efficiency of PVT and grid cellular networks is investigated. When the path loss exponent is fixed, numerical results of PVT and grid cellular networks consistently demonstrate that the spatial spectrum efficiency decreases with the increase of the call arrival rate. When the call arrival rate increases, the blocking probability in the cell is correspondingly increased. The increase of blocking probability in the cell in turn reduces the spatial spectrum efficiency of PVT and grid cellular networks.

\begin{figure}
\centerline{\includegraphics[width=7.5cm,draft=false]{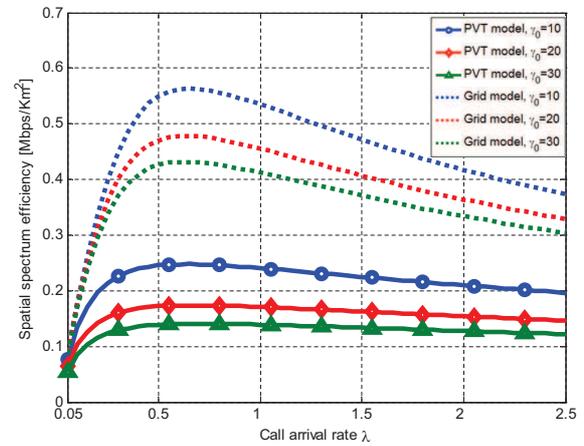}}
\caption{\small Spatial spectrum efficiency with respect to call arrival rate $\lambda$ considering different SINR thresholds $\gamma_0$ in PVT and grid cellular networks.}
\label{fig8}
\end{figure}

Fig.~\ref{fig8} shows the spatial spectrum efficiency with respect to the call arrival rate considering different SINR thresholds in PVT and grid cellular networks. When the call arrival rate in cellular scenarios is fixed, the spatial spectrum efficiency decreases with the increase of the SINR threshold. When the SINR threshold is fixed, there exist thresholds for different call arrival rates in cellular scenarios. Below the threshold, the spatial spectrum efficiency increases with the increase in the call arrival rate and above the threshold the spatial spectrum efficiency decreases with the increase in the call arrival rate. Numerical results of PVT and grid cellular networks consistently validate that there exist maximum spatial spectrum efficiency values considering different call arrival rates in cellular scenarios.

\begin{figure}
\centerline{\includegraphics[width=7.5cm,draft=false]{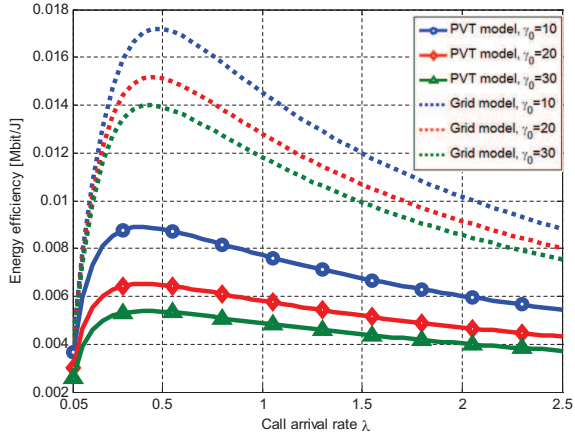}}
\caption{\small Energy efficiency with respect to call arrival rate $\lambda$ considering different SINR thresholds $\gamma_0$ in PVT and grid cellular networks.}
\label{fig9}
\end{figure}

The impact of the call arrival rate on the energy efficiency of PVT and grid random cellular networks is evaluated in Fig.~\ref{fig9}. When the SINR threshold is fixed, there exist thresholds for different call arrive rates in cellular scenarios. Below the threshold, the energy efficiency increases and above the threshold the energy efficiency decreases with the increase in the call arrival rate. Numerical results of PVT and grid cellular networks consistently validate that there exist maximum energy efficiency values considering different call arrive rates in cellular scenarios. Therefore, to achieve an optimal spectrum and energy efficiency of cellular networks, the call arrive rate and the SINR threshold in a cell should be considered carefully by telecommunications operators.

\begin{figure}
\centerline{\includegraphics[width=7.5cm,draft=false]{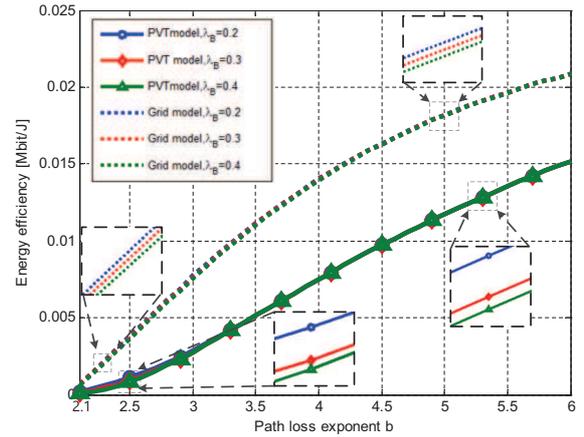}}
\caption{\small Energy efficiency with respect to path loss exponent $b$ considering different BS densities $\lambda_B$ in PVT and grid cellular networks.}
\label{fig10}
\end{figure}

Fig.~\ref{fig10} depicts the energy efficiency with respect to the path loss exponent considering different BS densities in PVT and grid cellular networks. When the BS density of cellular networks is fixed, numerical results of PVT and grid cellular networks consistently confirm that the energy efficiency increases with the increase of the path loss exponent. With the increase of the path loss exponent, both the desired signal and the interference are exponentially attenuated over wireless channels. Since the distance between the interfering transmitters and the receiver is longer than the distance between the desired BS and the receiver, the interference will experience larger attenuation than the desired signal when the path loss exponent increases. Therefore, the outage probability decreases with the increase of the path loss exponent. This result implies that the energy and spatial spectrum efficiency increase with the increase of the path loss exponent in Fig.~\ref{fig6}, Fig.~\ref{fig7} and Fig.~\ref{fig10}. When the path loss exponent is fixed, the energy efficiency decreases with the increase of the BS density in PVT and grid cellular networks.

\section{Conclusions}
\label{sec5}
To evaluate the spatial spectrum and energy efficiency in network level, the Markov chain is first integrated into the PVT random cellular networks in this paper. Based on the Markov chain based channel access model, spatial spectrum and energy efficiency are analyzed for PVT random cellular networks. To derive these models, a Markov chain is first presented for modeling of wireless channel access in a typical PVT cell. Moreover, taking into account the path loss and Rayleigh fading effects over wireless channels, the outage probability and the blocking probability are derived for a typical PVT cell. Furthermore, the spatial spectrum and energy efficiency are obtained for PVT random cellular networks. Numerical results have shown that the call arrival rate in a PVT cell and the BS density of PVT random cellular networks have adverse effects on the spatial spectrum efficiency of PVT random cellular networks. Moreover, the path loss exponent and the SINR threshold have great impact on the energy efficiency of PVT cellular networks. In the end, our results provide insights into the evaluation of spatial spectrum and energy efficiency of PVT random cellular networks considering different call arrival rates in cellular scenarios.

\begin{IEEEbiography}[{\includegraphics[width=1in,height=1.25in,clip,keepaspectratio]{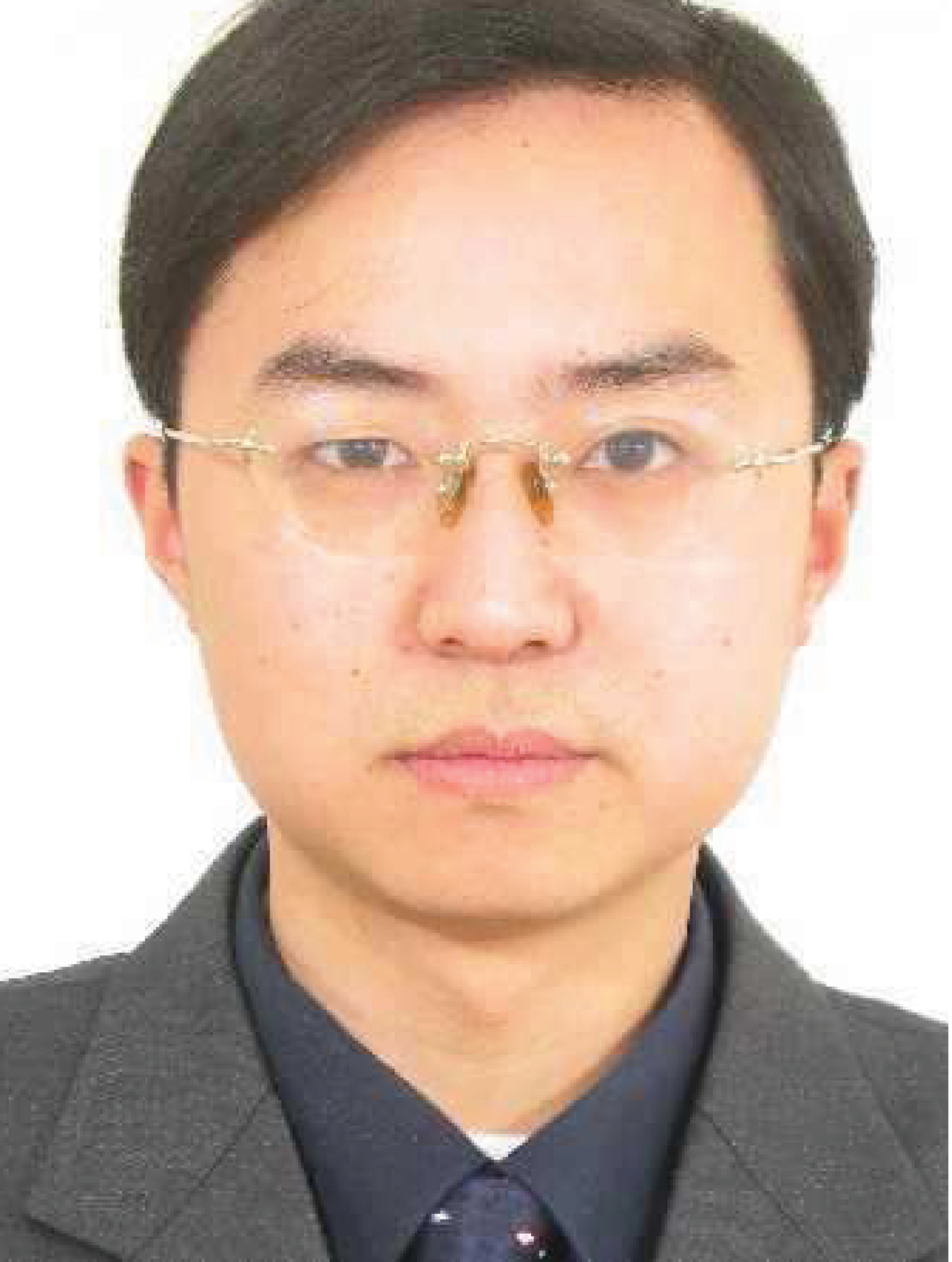}}]{Xiaohu~Ge}
(M'09-SM'11) is currently a Professor with the School of Electronic Information and Communications at Huazhong University of Science and Technology (HUST), China. He received his PhD degree in Communication and Information Engineering from HUST in 2003. He has worked at HUST since Nov. 2005. Prior to that, he worked as a researcher at Ajou University (Korea) and Politecnico Di Torino (Italy) from Jan. 2004 to Oct. 2005. He was a visiting researcher at Heriot-Watt University, Edinburgh, UK from June to August 2010. His research interests are in the area of mobile communications, traffic modeling in wireless networks, green communications, and interference modeling in wireless communications. He has published about 90 papers in refereed journals and conference proceedings and has been granted about 15 patents in China. He received the Best Paper Awards from IEEE Globecom 2010. He is leading several projects funded by NSFC, China MOST, and industries. He is taking part in several international joint projects, such as the EU FP7-PEOPLE-IRSES: project acronym S2EuNet (grant no. 247083), project acronym WiNDOW (grant no. 318992) and project acronym CROWN (grant no. 610524).

Dr. Ge is a Senior Member of the China Institute of Communications and a member of the National Natural Science Foundation of China and the Chinese Ministry of Science and Technology Peer Review College. He has been actively involved in organizing more the ten international conferences since 2005. He served as the Executive Chair for the 2013 IEEE International Conference on Green Computing and Communications (IEEE GreenCom) and as the Cochair of the Workshop on Green Communication of Cellular Networks at the 2010 IEEE GreenCom. He serves as an Associate Editor for the \textit{IEEE ACCESS}, \textit{Wireless Communications and Mobile Computing Journal (Wiley)} and \textit{the International Journal of Communication Systems (Wiley)}, etc.
Moreover, he served as the guest editor for \textit{IEEE Communications Magazine} Special Issue on 5G Wireless Communication Systems and \textit{ACM/SpringMobile Communications and Application} Special Issue on Networking in 5G Mobile Communication Systems.
\end{IEEEbiography}

\begin{IEEEbiography}[{\includegraphics[width=1in,height=1.25in,clip,keepaspectratio]{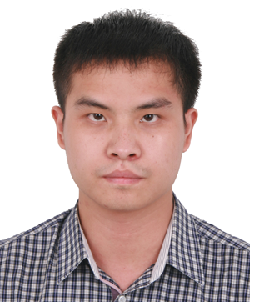}}]{Bin~Yang}
(S'13) received the Bachelor¡¯s degree in communication and information system from Huazhong University of Science and Technology, Wuhan, China, in 2012, where he is currently working toward the Doctorate degree.

His research interests include queuing theory, stochastic geometry, and heterogeneous networks.
\end{IEEEbiography}

\begin{IEEEbiography}[{\includegraphics[width=1in,height=1.25in,clip,keepaspectratio]{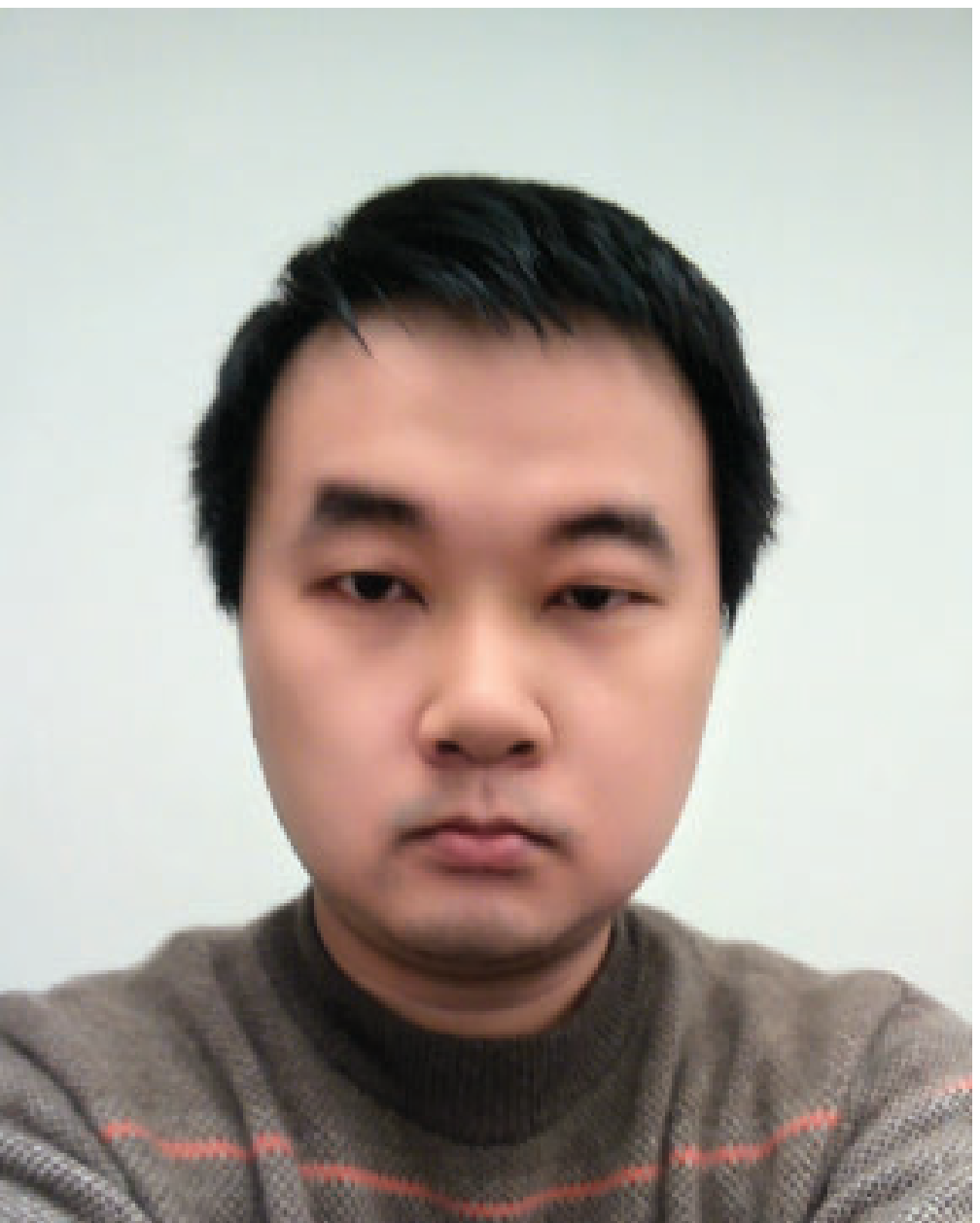}}]{Junliang~Ye}
recieved the B.Sc. Degree in communication engineering from China University of Geosciences, Wuhan, P.R China, in 2011, and he is currently a Ph.D student in communication and information system in Huazhong University of Science and Technology, Wuhan, P.R China.

His research interests include heterogeneous networks, stochastic geometry, mobility based access models of cellular networks and next generation wireless communication.
\end{IEEEbiography}

\begin{IEEEbiography}[{\includegraphics[width=1in,height=1.25in,clip,keepaspectratio]{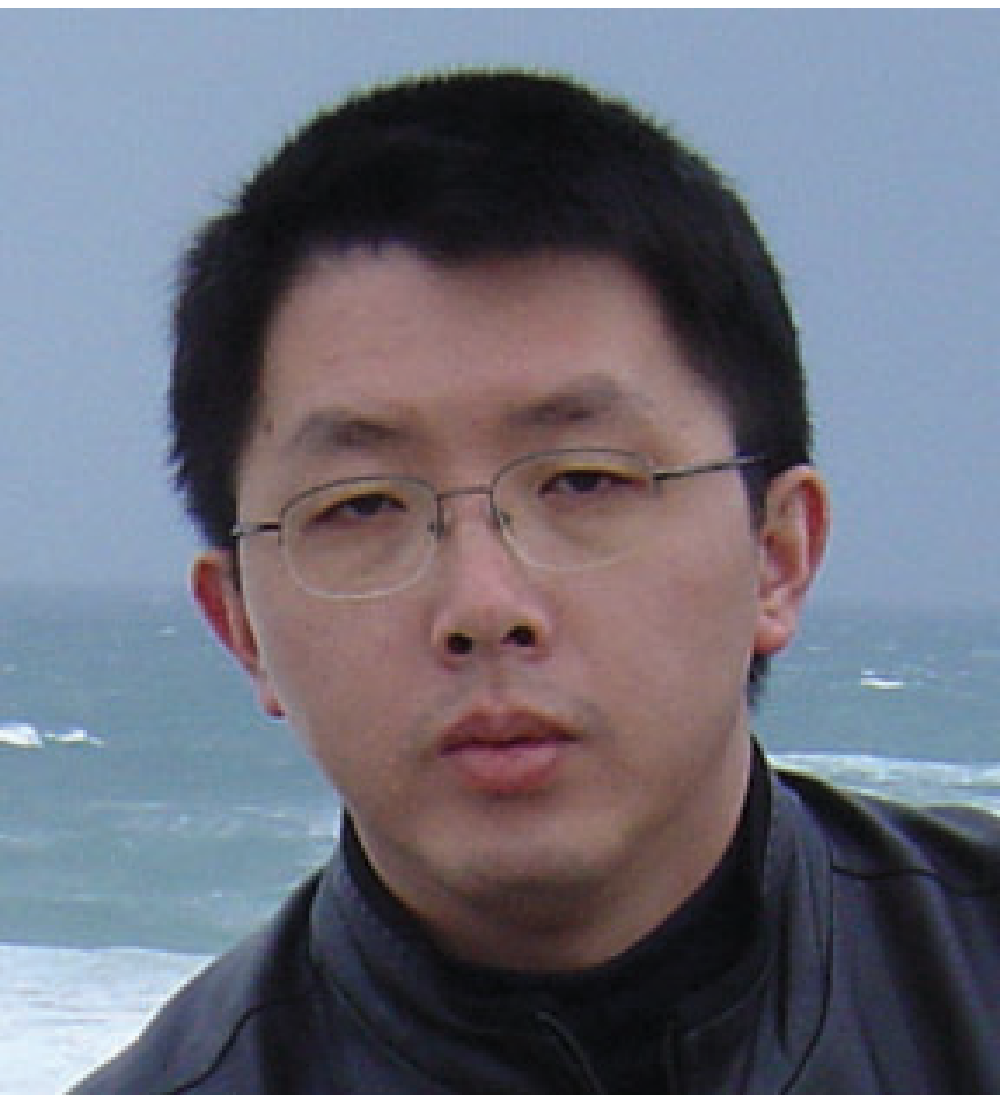}}]{Guoqiang~Mao}
(S'98-M'02-SM'08) received PhD in telecommunications engineering in 2002 from Edith Cowan University. He currently holds the position of Professor of Wireless Networking, Director of Center for Real-time Information Networks at the University of Technology, Sydney. He has published more than 100 papers in international conferences and journals, which have been cited more than 3000 times.

His research interest includes intelligent transport systems, applied graph theory and its applications in telecommunications, wireless sensor networks, wireless localization techniques and network performance analysis.
\end{IEEEbiography}

\vspace{-5 mm}

\begin{IEEEbiography}[{\includegraphics[width=1in,height=1.25in,clip,keepaspectratio]{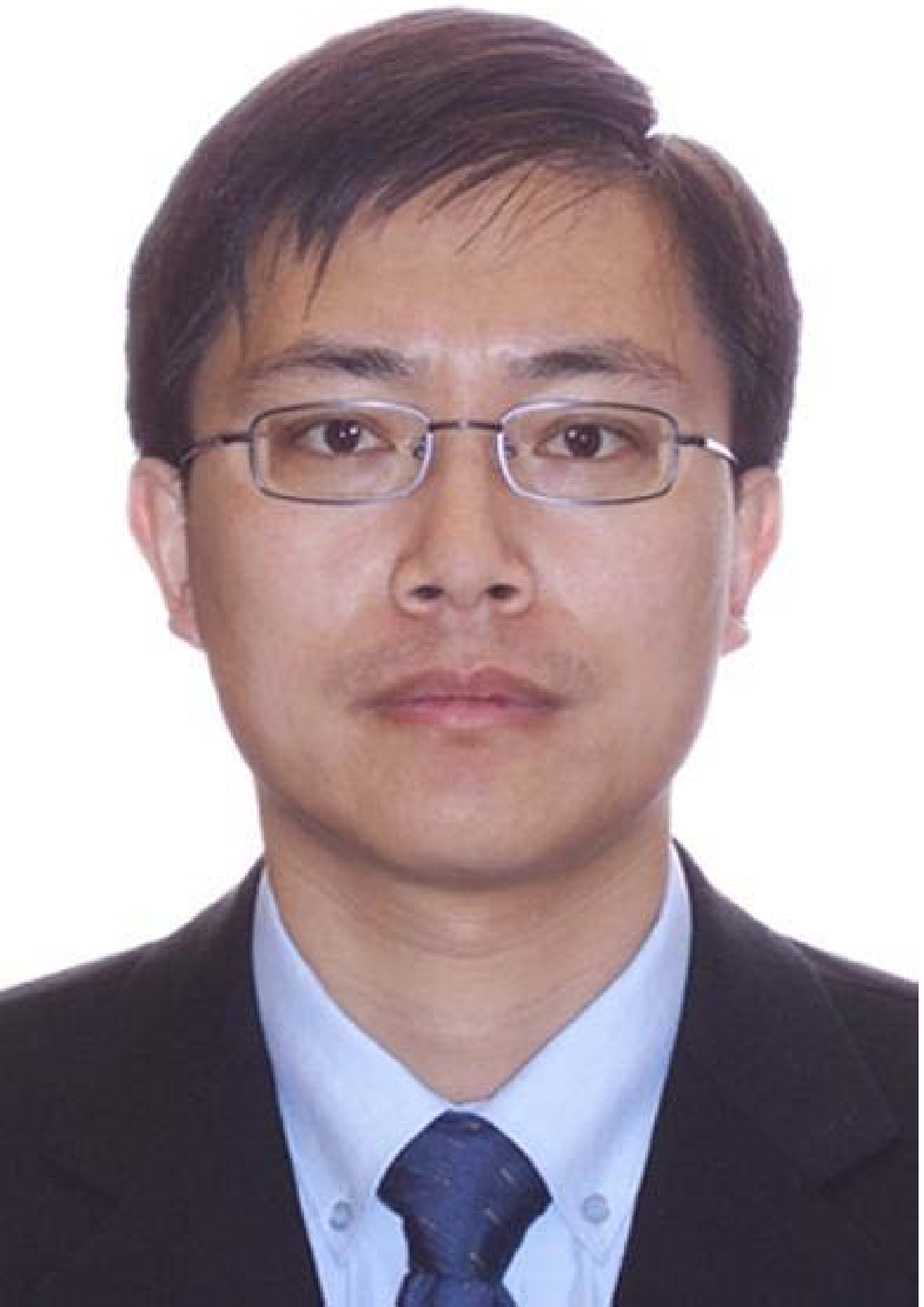}}]{Cheng-Xiang~Wang}
(S'01-M'05-SM'08) received the BSc and MEng degrees in Communication and Information Systems from Shandong University, China, in 1997 and 2000, respectively, and the PhD degree in Wireless Communications from Aalborg University, Denmark, in 2004.

He has been with Heriot-Watt University, Edinburgh, U.K., since 2005 and was promoted to a Professor in 2011. He is also an Honorary Fellow of the University of Edinburgh, U.K., and a Chair/Guest Professor of Shandong University and Southeast University, China. He was a Research Fellow at the University of Agder, Grimstad, Norway, from 2001-2005, a Visiting Researcher at Siemens AG Mobile Phones, Munich, Germany, in 2004, and a Research Assistant at Technical University of Hamburg-Harburg, Hamburg, Germany, from 2000--2001. His current research interests include wireless channel modeling, green communications, cognitive radio networks, vehicular communication networks, massive MIMO, millimeter wave communications, and 5G wireless communication networks. He is the Editor of one book. He has published 1 book chapter and over 210 papers in refereed journals and conference proceedings.

Prof. Wang served or is currently serving as an editor for 8 international journals, including IEEE Transactions on Vehicular Technology (2011--) and IEEE Transactions on Wireless Communications (2007--2009). He was the leading Guest Editor for IEEE Journal on Selected Areas in Communications, Special Issue on Vehicular Communications and Networks. He served or is serving as a TPC member, TPC Chair, and General Chair for over 70 international conferences. He received the Best Paper Awards from IEEE Globecom 2010, IEEE ICCT 2011, ITST 2012, and IEEE VTC 2013-Fall. He is a Fellow of the IET, a Fellow of the HEA, and a member of EPSRC Peer Review College.
\end{IEEEbiography}

\vspace{-6 mm}

\begin{IEEEbiography}[{\includegraphics[width=1in,height=1.25in,clip,keepaspectratio]{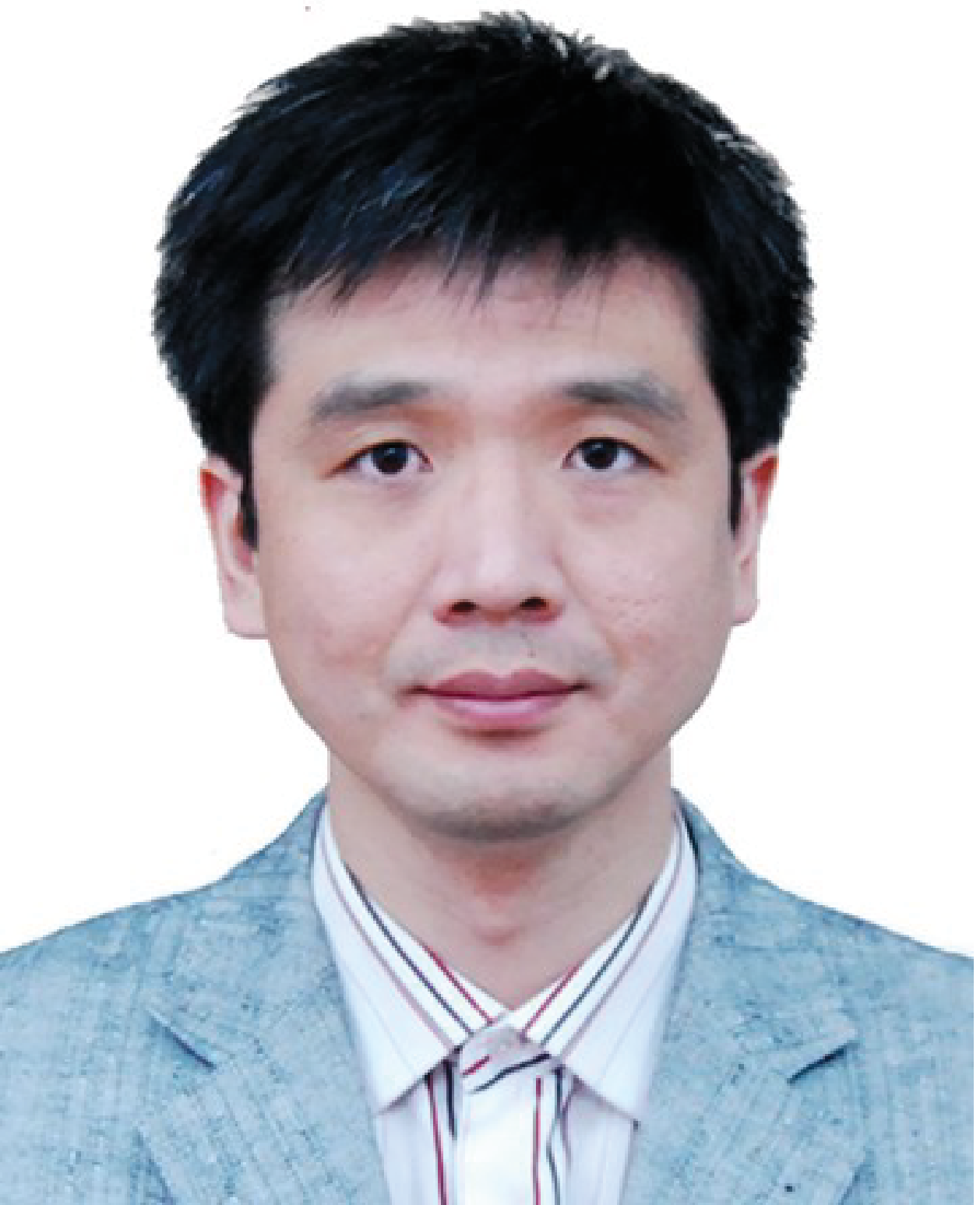}}]{Tao~Han}
(M'13) received the Ph.D. degree in communication and information engineering from Huazhong University of Science and Technology (HUST), Wuhan, China in December, 2001.

He is currently an Associate Professor with the School of Electronic Information and Communications, HUST. From August, 2010 to August, 2011, he was a Visiting Scholar with University of Florida, Gainesville, FL, USA, as a Courtesy Associate Professor. His research interests include wireless communications, multimedia communications, and computer networks.

Dr. Han is currently serving as an Area Editor for the \emph{EAI Endorsed Transactions on Cognitive Communications}.
\end{IEEEbiography}

\end{document}